\documentclass[12pt]{article}
\usepackage{amssymb,epsf,euscript,graphicx}
\usepackage{subfigure}

 \usepackage[margin=1.0in]{geometry}

 \usepackage{wrapfig}

 \usepackage[hyphens]{url} 
\usepackage{hyperref}  
\hypersetup{breaklinks=true}  

\usepackage{cite}

\usepackage{enumitem}

\usepackage{lipsum} 

\usepackage{mathrsfs}
\usepackage{amsmath} 

\newcommand{\beq}{\begin{equation}}
\newcommand{\eeq}{\end{equation}}
\newcommand{\barr}{\begin{array}}
\newcommand{\earr}{\end{array}}
\newcommand{\beqarr}{\begin{eqnarray}}
\newcommand{\eeqarr}{\end{eqnarray}}
\newcommand{\beqar}{\begin{eqnarray*}}
\newcommand{\eeqar}{\end{eqnarray*}}
\newcommand{\bef}{\begin{figure}}
\newcommand{\eef}{\end{figure}}

\newcommand{\bm}[1]{\mbox{\boldmath$#1$}}

    \usepackage{fancyhdr}
\usepackage{blindtext}
\fancypagestyle{mypagestyle}{
\fancyhf{}
\fancyfoot[C]{\thepage}  

}
\makeatletter
\let\ps@plain\ps@mypagestyle
\makeatother
\fancypagestyle{firstpage}
{
\chead{
\makebox[\linewidth]{\rule{\paperwidth}{2.0pt}}
\textbf{PROJECT DESCRIPTION}
\vspace{-0.2in}
\makebox[\linewidth]{\rule{\paperwidth}{2.0pt}}
}
}

\fancypagestyle{lastpage}
{
\chead{
\makebox[\linewidth]{\rule{\paperwidth}{2.0pt}}
\textbf{BIBLIOGRAPHY}
\vspace{-0.2in}
\makebox[\linewidth]{\rule{\paperwidth}{2.0pt}}
}
}

\begin{document}

\begin{center}
\textbf{\large{On the Mathematical Analysis and Physical Implications of the Principle of Minimum Pressure Gradient}}\\
By Haithem E. Taha
\end{center}

\begin{abstract}
In this paper, we establish a two-way equivalence between the incompressible Navier-Stokes equation (INSE) and the principle of minimum pressure gradient (PMPG). We prove that a candidate smooth flow field is a solution of the INSE \textit{if and only if} its instantaneous evolution minimizes, at every instant, the norm of the pressure force, required to enforce incompressibility. We show that the PMPG is precisely the minimization formulation of the Leray-Helmholtz projection. Any admissible instantaneous evolution (e.g., onset of separation) resulting from the INSE necessarily minimizes the PMPG cost. Conversely, any other kinematically admissible evolution, requiring a strictly larger pressure force to ensure the same constraints, does not satisfy the INSE. Thus, the PMPG offers a variational perspective through which intricate incompressible flow behaviors may be interpreted.

In a finite-dimensional setting with divergence-free modes, we show that the PMPG yields the same dynamics as classical Galerkin projection. Moreover, the PMPG provides a natural generalization of classical Galerkin projection beyond linear modal expansions, accommodating nonlinear and non-modal representations. We then examine the relation between instantaneous dynamical minimization and steady variational selection, including its connection to the variational theory of lift. Motivated by these observations, we formulate conjectures concerning necessary conditions for stability and the convergence of Navier–Stokes solutions to Euler's in the vanishing-viscosity limit.
\end{abstract}


\section{Introduction}
Gauss's principle of least constraint transforms a constrained dynamics problem into pure minimization. It asserts that a mechanical system evolves from one time instant to another in the closest possible manner to its \textit{free motion}---the motion that would occur in the absence of constraints \cite{Gauss_Least_Constraint,Gauss_Vs_Least_Action,Papastavridis,Udwadia_Kalaba_Book}. More precisely, among all kinematically admissible motions, the actual motion is the one for which the deviation from the free motion is minimal at each instant. This deviation is directly proportional to the constraint forces, required to ensure the kinematical and geometrical constraints. In this sense, Gauss's principle asserts that a mechanical system evolves by minimizing the magnitude of the constraint force. Any alternative evolution would require an unnecessarily larger constraint force to maintain the same constraints.

It can be shown that the first-order necessary condition of optimality for the Gaussian cost is Newton's equations of motion \cite{NS_QP_IEEE}. That is, If the acceleration of a trajectory minimizes the Gaussian cost at every instant, among all kinematically admissible accelerations, then the trajectory necessarily satisfies Newton's equations of motion. This result can be strengthened further for mechanical systems subject to smooth forcing functions, for which solutions of Newton's equations are unique. In this case, a candidate admissible trajectory is a solution of Newton's equations of motion \textit{if and only if} its acceleration minimizes the Gaussian cost at every instant.

In this paper, we demonstrate this minimization framework of Gauss's principle using the double-pendulum as a concrete example. At a given instant, for which the two angles $(\theta,\phi)$ and angular velocities $(\dot\theta,\dot\phi)$ are prescribed, there exist infinitely many possible evolutions; i.e. accelerations $(\ddot\theta,\ddot\phi)$; all of which are kinematically admissible. Each such evolution, however, requires a different magnitude of constraint force to maintain the pendulum constraints. In other words, on the two-dimensional configuration manifold (the torus in this case), there are infinitely many admissible directions in the tangent bundle, yet only one evolution occurs in reality. It is the one that satisfies Newton's equations of motion. Equivalently, according to Gauss's principle, it is the evolution that requires the smallest possible constraint force to maintain the pendulum constraints. Any alternative evolution would necessarily require a strictly larger constraint force to ensure the same constraints.

In the absence of impressed forced (e.g., pendulum in the horizontal plane), this minimization framework implies motion along \textit{geodesics} (i.e., straight lines) on the configuration manifold. That is, the trajectories of motion are the least curved paths on the configuration manifold, following Hertz's principle of least curvature \cite{Papastavridis}. If the system is subject to impressed forces, then Gauss's principle becomes the minimization framework of projecting the impressed forces onto the configuration manifold.

Gauss's principle was recently extended from particle mechanics to continuum mechanics of incompressible flows, leading to the principle of minimum pressure gradient (PMPG) \cite{PMPG_PoF}. Recognizing that, for incompressible flows, the pressure force plays the role of a constraint force that maintains the continuity constraint, the PMPG asserts that an incompressible flow evolves from one instant to the next by minimizing the $\mathcal{L}^2$-norm of the pressure force, required to ensure incompressibility. Similar to the fact that Newton's equations of motion arise as the first-order necessary condition of optimality for the Gaussian cost in particle mechanics, it was shown that the Navier-Stokes equation is the first-order necessary condition of optimality for the PMPG cost. That is, if the local acceleration of an incompressible flow minimizes the $\mathcal{L}^2$-norm of the pressure force at every instant, among all divergence-free accelerations, then the resulting flow field necessarily satisfies the Navier-Stokes equation.

In this paper, we strengthen this result by proving that a candidate smooth flow field is a solution of the Navier-Stokes equation \textit{if and only if} its local acceleration minimizes the $\mathcal{L}^2$-norm of the pressure force at every instant. This mathematical fact establishes a two-way equivalence between the Navier-Stokes equation (NSE) and the PMPG minimization framework. In particular, if the acceleration of a flow field minimizes the PMPG cost at every instant, then the flow necessarily satisfies the NSE; conversely, if at any instant the local acceleration requires a larger pressure-gradient force than another kinematically admissible evolution, then the corresponding flow field cannot be a solution of the NSE.

This two-way equivalence suggests that the PMPG may be viewed as a causal mechanism for incompressible flow phenomena. If a smooth incompressible flow field behaves a certain way, we may attribute such a behavior to the minimization of the pressure gradient, since any alternative behavior would require an unnecessarily larger pressure force to maintain incompressibility and therefore, by virtue of the two-way equivalence, could not be a solution of the NSE.

Noting that any orthogonal projection can be formulated as a minimization problem, we show that the PMPG provides the minimization framework underlying the Leray projection: projecting the NSE onto the space of divergence-free fields \cite{Chorin_Projection,Temam_Projection,Moin_Incompressible1,Pressure_BCs,Chorin_Marsden_Book}. We also clarify several common points of confusion arising from conflating the PMPG with the principle of stationary action or with the Dirichlet principle associated with the pressure Poisson equation in projection methods. We conclude by posing conjectures that may be of interest for further mathematical analysis, including an inviscid stability criterion and the convergence of Navier-Stokes solutions to Euler's.

\section{Gauss's Principle of Least Constraint}\label{Sec:Gauss}
\begin{wrapfigure}{l}{0.50\textwidth}
\vspace{-0.15in}
 \begin{center}
 \includegraphics[width=7cm]{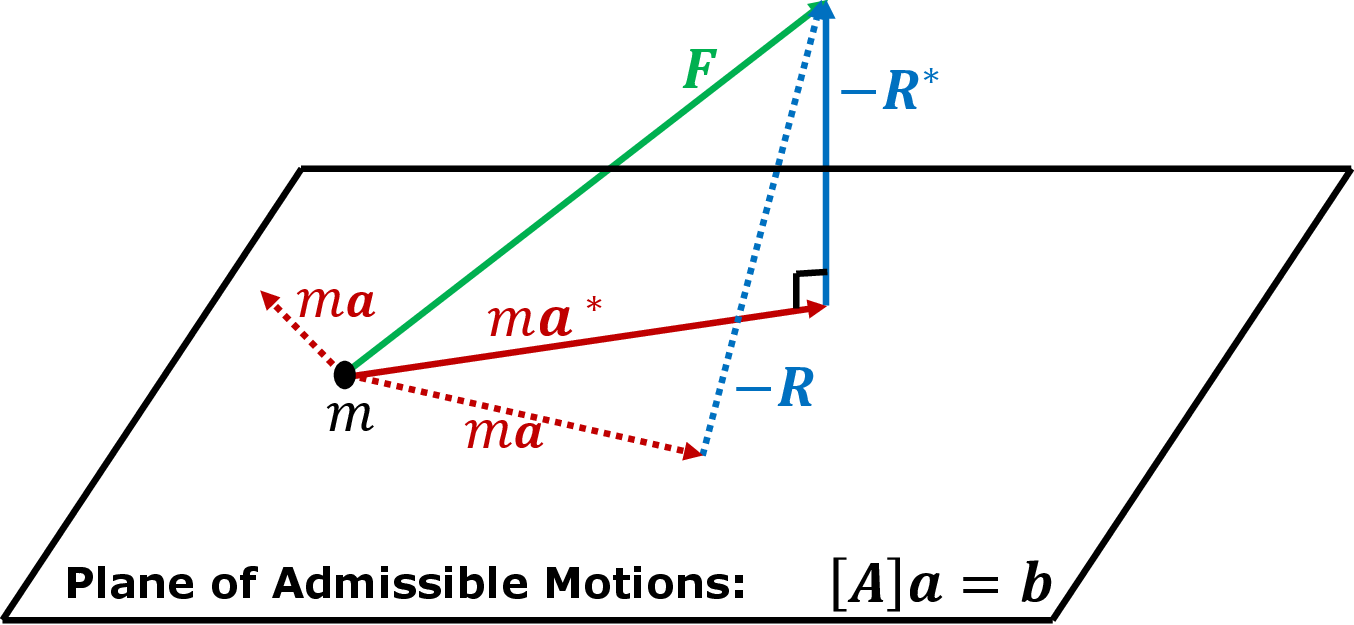}  \vspace{-0.15in}
 \caption{Schematic illustration of the instantaneous plane of admissible motion, defined by (\ref{eq:Constraints_Linear}), at a particular configuration. Infinitely many instantaneous motions lie this plane (dotted vectors), each satisfying the linear constraint (\ref{eq:Constraints_Linear}) at the expense of a constraint force $\bm{R}$.  Gauss's principle asserts that, among all kinematically admissible motions, Nature selects the one that requires the least constraint force. This minimization is simply equivalent to projecting the impressed force $\bm{F}$ onto the plane of admissible motions.}\vspace{-0.15in}
 \label{Fig:Gauss_Schematic} 
 \end{center}\vspace{-0.15in}
\end{wrapfigure}

In his four-page philosophical note, published in a journal that still exists today, Gauss \cite{Gauss_Least_Constraint} postulated one of the fundamental principles of mechanics. Consider a particle of mass $m$ subject to an \textit{impressed} force $\bm{F}$, as illustrated in Fig.~\ref{Fig:Gauss_Schematic}. In the absence of constraints, it would accelerate in the direction of the force with acceleration $\bm{a}^{\rm{free}} = \frac{\bm{F}}{m}$. Gauss referred to this motion as the \textit{free motion}---the motion that would occur in the absence of constraints. However, if the particle is constrained to accelerate within some instantaneous plane of admissible motion defined by the constraint $[\bm{A}]\bm{a} = \bm{b}$, then the actual motion must necessarily deviate from the free motion. Since this deviation arises solely due to the constraint, Gauss’s profound insight was that it must be the \textit{least} deviation compatible with the constraint. Nature will not overdo it. He wrote:
\begin{center}
``\textit{The motion of a system of $N$ material points takes place in every moment in maximum accordance with the free movement or under least constraint, the measure of constraint, is considered as the sum of products of mass and the square of the deviation to the free motion.}"
\end{center}

 \noindent Gauss's principle admits an equally illuminating interpretation. On the instantaneous admissible plane of motion, there exist infinitely many candidate accelerations, as illustrated in Fig.~\ref{Fig:Gauss_Schematic}. Each candidate requires a specific \textit{constraint} force $\bm{R}$ such that, when combined with the impressed force $\bm{F}$, the resulting motion
\[ m\bm{a}=\bm{F}+\bm{R}\]
satisfies the constraint. The force $\bm{R}$ exists solely to enforce the constraint; its \textit{raison d'être} is the constraint itself—if the constraint is removed, $\bm{R}$ vanishes. Gauss then asserted that Nature selects the motion requiring the smallest possible magnitude of the constraint force necessary to enforce the constraint, hence the term \textit{Least Constraint}. Any alternative candidate would demand an unnecessarily larger constraint force to ensure the same constraint, and is therefore not realized.

Jacobi \cite{Jacobi} later gave Gauss's principle an explicit mathematical formulation by introducing the quadratic cost function:
\begin{equation}\label{eq:Gauss}
  Z = \frac{1}{2} \sum_{i=1}^N m_i \left| \bm{a}_i- \frac{\bm{F}_i}{m_i} \right|^2,
\end{equation}
where $N$ denotes the number of particles and $\bm{a}_i$ is the inertial acceleration of the $i^{\rm th}$ particle. According to Gauss' principle, $Z$ must be minimum at every instant, provided that the constraints are satisfied. Assume the particles evolve in a $d$-dimensional space so that the stacked acceleration vector $\bm{a}=[\bm{a}_1^T, ..., \bm{a}_N^T]^T\in\mathbb{R}^{dN}$ collects all inertial accelerations. Suppose the system is subject to $c\le dN$ constraints, possibly nonlinear in positions and velocities. Differentiation of the constraint equations with respect to time yields a relation linear in the accelerations:
\begin{equation}\label{eq:Constraints_Linear}
A_{\ell j} \; a_j = b_\ell
\end{equation}
for some $\bm{A}\in\mathbb{R}^{c\times dN}$, $\bm{b}\in\mathbb{R}^c$.

Under Newtonian mechanics, solving the dynamics requires $dN+c$ equations in $dN+c$ unknowns. One has the $dN$ equations of motion
\begin{equation}\label{eq:Newton}
  m_i \bm{a}_i = \bm{F}_i + \bm{R}_i \;\;\; \forall i=1,..,N,
\end{equation}
together with the $c$ constraint equations \eqref{eq:Constraints_Linear}. The unknowns are the $dN$ accelerations $\bm{a}$ and the $c$ independent components of the constraint forces $\bm{R}_i$.

Gauss's principle transforms this dynamics problem into the following minimization problem:
\begin{equation}\label{eq:Gauss_QP_Problem}
\min_{\bm{a}} \; Z(\bm{a}) \;\;\; \mbox{s.t.} \;\;\; [\bm{A}(\bm{x},\bm{v})] \; \bm{a} = \bm{b},
\end{equation}
In our recent effort \cite{NS_QP_IEEE}, we showed that this minimization problem is a strongly convex quadratic programming problem and therefore admits a unique solution. Moreover, its first-order necessary conditions of optimality coincide precisely with Newton's equations of motion. Thus, the unique solution of Gauss's minimization problem (\ref{eq:Gauss_QP_Problem}) necessarily satisfies Newton's equation of motion.

\begin{wrapfigure}{l}{0.4\textwidth}
\vspace{-0.1in}
 \begin{center}
 \includegraphics[width=5cm]{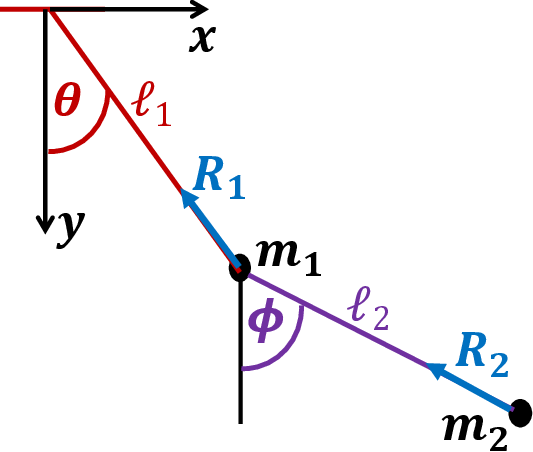}  \vspace{-0.1in}
 \caption{A Schematic for a double-pendulum oscillating in a horizontal plane.}
 \label{Fig:Double_Pendulum_Schematic} \vspace{-0.1in}
 \end{center}
\end{wrapfigure} \noindent \textbf{Double Pendulum Example:} Because Gauss's principle is rarely covered in engineering graduate curricula, it is instructive to illustrate it through a simple example that also clarifies its minimization structure. Consider a double pendulum constrained to move in a horizontal plane, as shown schematically in Fig.~\ref{Fig:Double_Pendulum_Schematic}. The masses $m_1$ and $m_2$ are constrained to remain at fixed distances $\ell_1$ and $\ell_2$ from their respective hinges. These geometric constraints give rise to constraint forces $R_1$ and $R_2$. In the absence of impressed forces, the \textit{free motion} (that would occur in the absence of the constraint) corresponds to zero acceleration. The Gaussian cost therefore reduces to
\begin{equation}\label{eq:Gaussian_Double_Pendulum}
Z = \frac{1}{2} \left[ m_1 \bm{a}_1^2 + m_2 \bm{a}_2^2\right].
\end{equation}

For simplicity, assume unit masses and unit lengths. The inertial accelerations $\bm{a}_1$ and $\bm{a}_2$ can then be expressed in terms of the generalized coordinates $\bm{q}=(\theta,\phi)$ as
\[ \bm{a}_1 = \ddot\theta \left(\begin{array}{c}\cos\theta\\-\sin\theta \end{array}\right) - \dot\theta^2 \left(\begin{array}{c}\sin\theta\\\cos\theta \end{array}\right), \;\; \bm{a}_2 = \bm{a}_1 + \ddot\phi \left(\begin{array}{c}\cos\phi\\-\sin\phi \end{array}\right) - \dot\phi^2 \left(\begin{array}{c}\sin\phi\\\cos\phi \end{array}\right). \]
Substituting $\bm{a}_1$ and $\bm{a}_2$ into the Gaussian cost \eqref{eq:Gaussian_Double_Pendulum}, one obtains a quadratic form in the accelerations $(\ddot\theta,\ddot\phi)$:
\begin{equation}\label{eq:Gaussian_Double_Pendulum_Quadratic}
Z = \frac{1}{2}\left[\begin{array}{cc} \ddot\theta & \ddot\phi \end{array}\right] \left[\begin{array}{cc} 2 & \cos(\theta-\phi) \\ \cos(\theta-\phi) & 1 \end{array}\right] \left(\begin{array}{c} \ddot\theta \\ \ddot\phi \end{array}\right) + \sin(\theta-\phi)  \left[\begin{array}{cc} \dot\phi^2 & -\dot\theta^2 \end{array}\right] \left(\begin{array}{c} \ddot\theta \\ \ddot\phi \end{array}\right) + h,
\end{equation}
where $h$ depends on $\bm{q}=(\theta,\phi)$ and $\dot{\bm{q}}=(\dot\theta,\dot\phi)$ but is independent of the accelerations $\ddot{\bm{q}}=(\ddot\theta,\ddot\phi)$.

The dynamics problem is therefore transformed, via Gauss's principle, to the minimization problem
\[ \min_{(\ddot\theta,\ddot\phi)} \; Z (\ddot\theta,\ddot\phi)\]
whose first-order necessary conditions for optimality are simply:
\[ \frac{\partial Z}{\partial \ddot\theta}=0 \;\; \mbox{and} \;\; \frac{\partial Z}{\partial \ddot\phi}=0, \]
yielding
\begin{equation}\label{eq:Double_Pendulum_EOM}
\left[\begin{array}{cc} 2 & \cos(\theta-\phi) \\ \cos(\theta-\phi) & 1 \end{array}\right] \left(\begin{array}{c} \ddot\theta \\ \ddot\phi \end{array}\right) =  \sin(\theta-\phi)  \left(\begin{array}{c} -\dot\phi^2 \\ \dot\theta^2 \end{array}\right),
\end{equation}
which coincide precisely with Newton's equations of motion for the double pendulum.

\begin{figure*}
\begin{center}
$\begin{array}{cc}
\subfigure[The plane of admissible accelerations.]{\label{Fig:Double_Pendulum_Tangent_Bundle}\includegraphics[width=7cm]{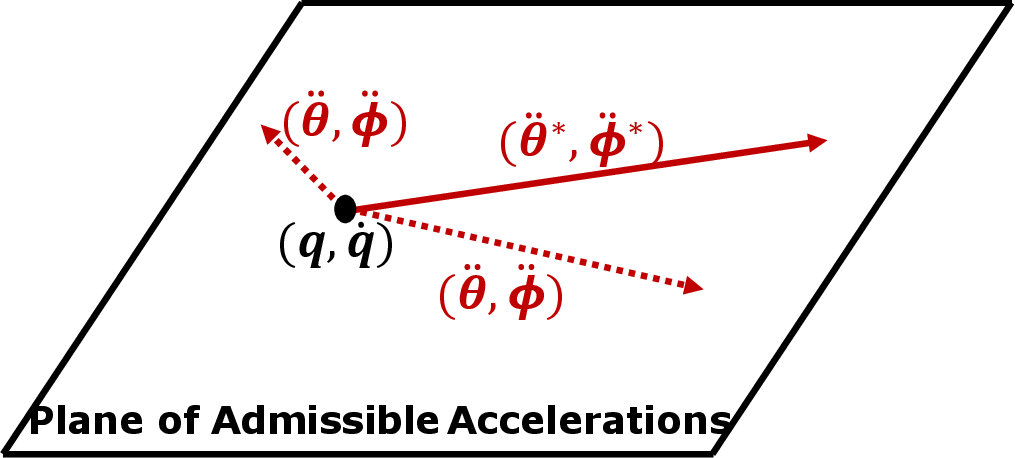}} & \subfigure[Variation of $Z$ with $\ddot\theta$, $\ddot\phi$.]{\label{Fig:Double_Pendulum_Z_Variation}\includegraphics[width=7cm]{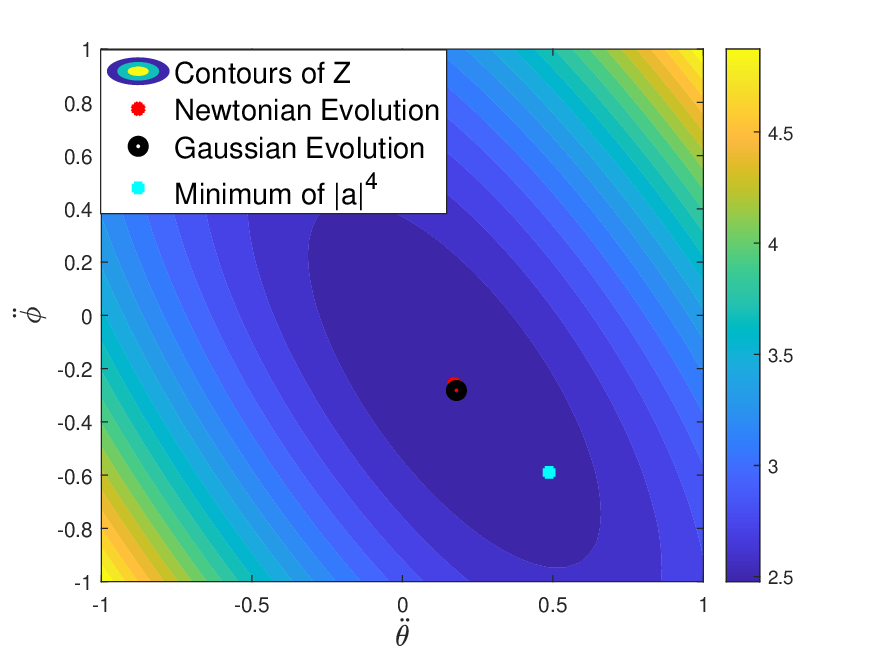}}
\end{array}$
\caption{A Schematic for the plane of admissible accelerations at a given point ($\bm{q},\dot{\bm{q}})$ in the tangent bundle, along with the contours of the Gaussian cost $Z$ in Eq. (\ref{eq:Gaussian_Double_Pendulum_Quadratic}) with the accelerations $\ddot\theta$, $\ddot\phi$ at the point $\bm{q}=(0^\circ,5^\circ)$, $\dot{\bm{q}}=(1,1)$. The minimizing accelerations $(\ddot\theta^*,\ddot\phi^*)$ satisfy Newton's equations of motion. However, the ones that minimize the 4-norm $\sum_i m_i \bm{a}_i^4$ do not necessarily coincide with the Newtonian evolution.}
\label{Fig:Double_Pendulum_Evolution}
\end{center}
\end{figure*}

It may be important to emphasize the insight gained from Gauss's principle as it pertains to the core discussion in this paper. While Gauss's principle and Newton's equations of motion are mathematically equivalent, the minimization formulation of Gauss's principe may provide additional insight. For example, at a given point ($\bm{q},\dot{\bm{q}})$ in the tangent bundle, there are infinitely many kinematically-admissible evolutions $(\ddot\theta,\ddot\phi)$, as shown in Fig. \ref{Fig:Double_Pendulum_Tangent_Bundle}. In the coordinate-system of $\bm{q}=(\theta,\phi)$, the pendulum constraints are automatically satisfied. For a given point ($\bm{q},\dot{\bm{q}})$ at a particular instant, there are infinitely many possible evolutions that respect the pendulum constraints. However, each kinematically-admissible evolution requires certain constraint forces $R_1$, $R_2$ to maintain tangency of the velocity vector to the configuration manifold.

Figure \ref{Fig:Double_Pendulum_Z_Variation} shows contours of the Gaussian cost $Z$ in Eq. (\ref{eq:Gaussian_Double_Pendulum_Quadratic}) in the acceleration space $(\ddot\theta,\ddot\phi)$ at the point $\bm{q}=(0^\circ,5^\circ)$, $\dot{\bm{q}}=(1,1)$. Each admissible evolution, represented by a candidate acceleration $(\ddot\theta,\ddot\phi)$, corresponds to a certain magnitude $Z$ of the constraint forces, required to maintain the pendulum constraints. The minimizing acceleration $(\ddot\theta^*,\ddot\phi^*)$ coincides precisely with that obtained from Newton's equations of motion. Any alternative candidate $(\ddot\theta,\ddot\phi)$ requires constraint forces of strictly larger magnitude.
Thus, Gauss's principle is not merely predictive; it also suggests a causal mechanism underlying the motion. From Gauss's perspective, the motion is driven by minimizing the magnitude of the constraint forces, required to maintain the constraints. Only this candidate evolution satisfies Newton's equations of motion. Any alternative evolution would demand an unnecessarily larger constraint force to ensure the same constraints, and would not satisfy Newton's equations.

In the absence of impressed forces, as in the horizontal-plane double pendulum, Gauss's principle yields least-curved trajectories. These are the closest trajectories to straight lines in the ambient Euclidean space; equivalently, they are geodesics on the configuration manifold itself \cite{Lanczos_Variational_Mechanics_Book,Bullo_Lewis_Book}. If the double pendulum is instead considered in the vertical plane, where it is subject to an impressed gravitational force, Gauss's principle reduces to the orthogonal projection of the impressed force onto the plane of admissible accelerations.

\section{The Principle of Minimum Pressure Gradient}
For incompressible flows, the Gaussian cost admits a natural extension to the continuum setting \cite{PMPG_PoF}:
\begin{equation}\label{eq:Gauss_Continuum}
  Z = \frac{1}{2} \int_\Omega \rho \left| \bm{a}- \bm{F} \right|^2 d\bm{x},
\end{equation}
where $\Omega$ denotes the spatial domain, $\bm{F}$ is the impressed force per unit mass, and $\bm{a}=\bm{u}_t + \bm{u}\cdot\bm\nabla \bm{u}$ is the total inertial acceleration of a fluid particle.

Incompressible flows are subject to pressure forces, viscous forces, and other body forces (e.g., gravity or electromagnetism), and are constrained to satisfy the continuity equation $\nabla\cdot\bm{u}=0$ together with the no-penetration boundary condition. As discussed in the preceding section, every constraint gives rise to a corresponding constraint force whose sole role is to enforce that constraint. For the continuity constraint, the Helmholtz decomposition reveals the underlying geometry of incompressible flows and the nature of the associated constraint force \cite{Chorin_Marsden_Book,Geometric_Control_Fluid_Dynamics,Helmholtz_Decomposition_Review}.

\begin{figure*}
\begin{center}
$\begin{array}{cc}
\subfigure[Schematic of the Helmholtz decomposition.]{\label{Fig:Helmholtz_Schematic}\includegraphics[width=7.5cm]{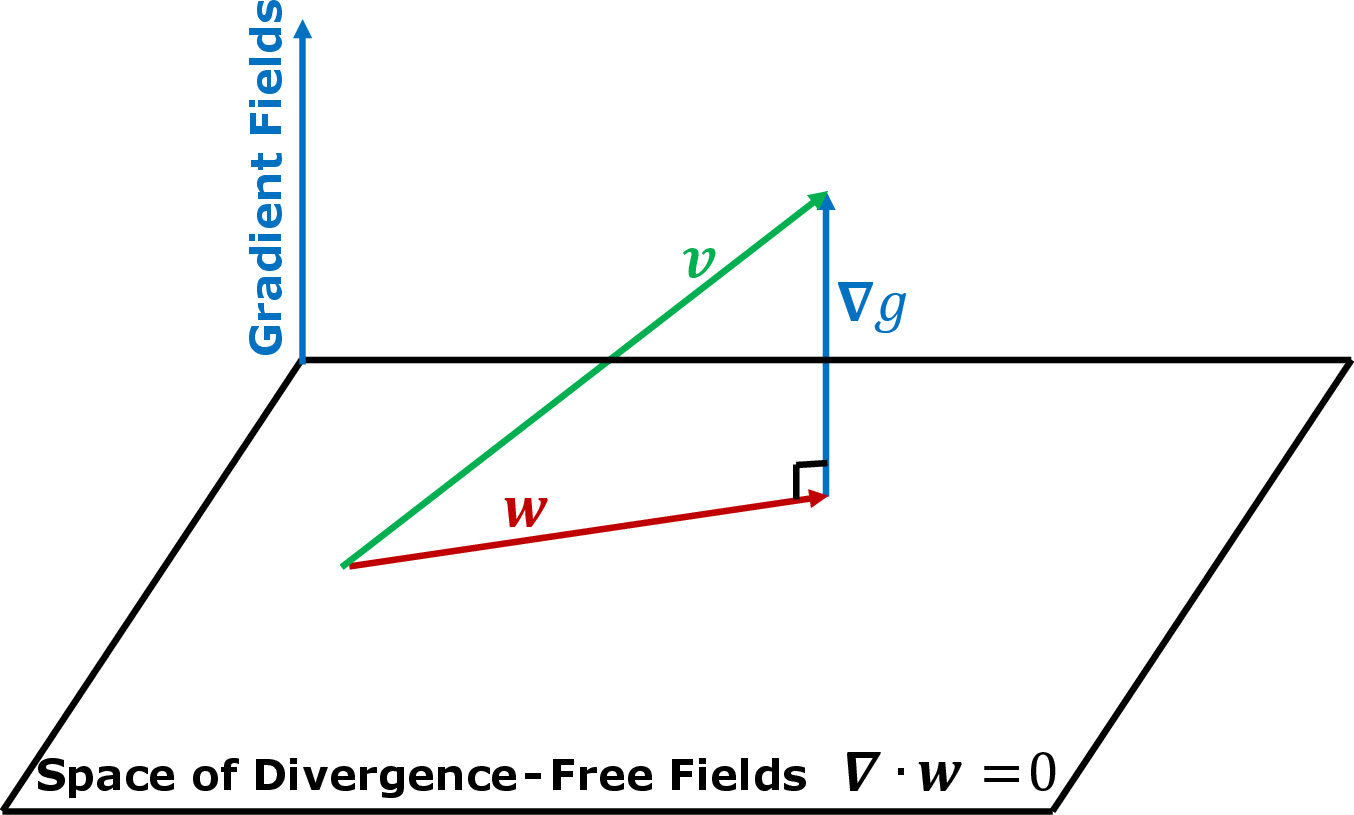}} & \subfigure[Geometry of incompressible flows.]{\label{Fig:Incompressible_Schematic}\includegraphics[width=7.5cm]{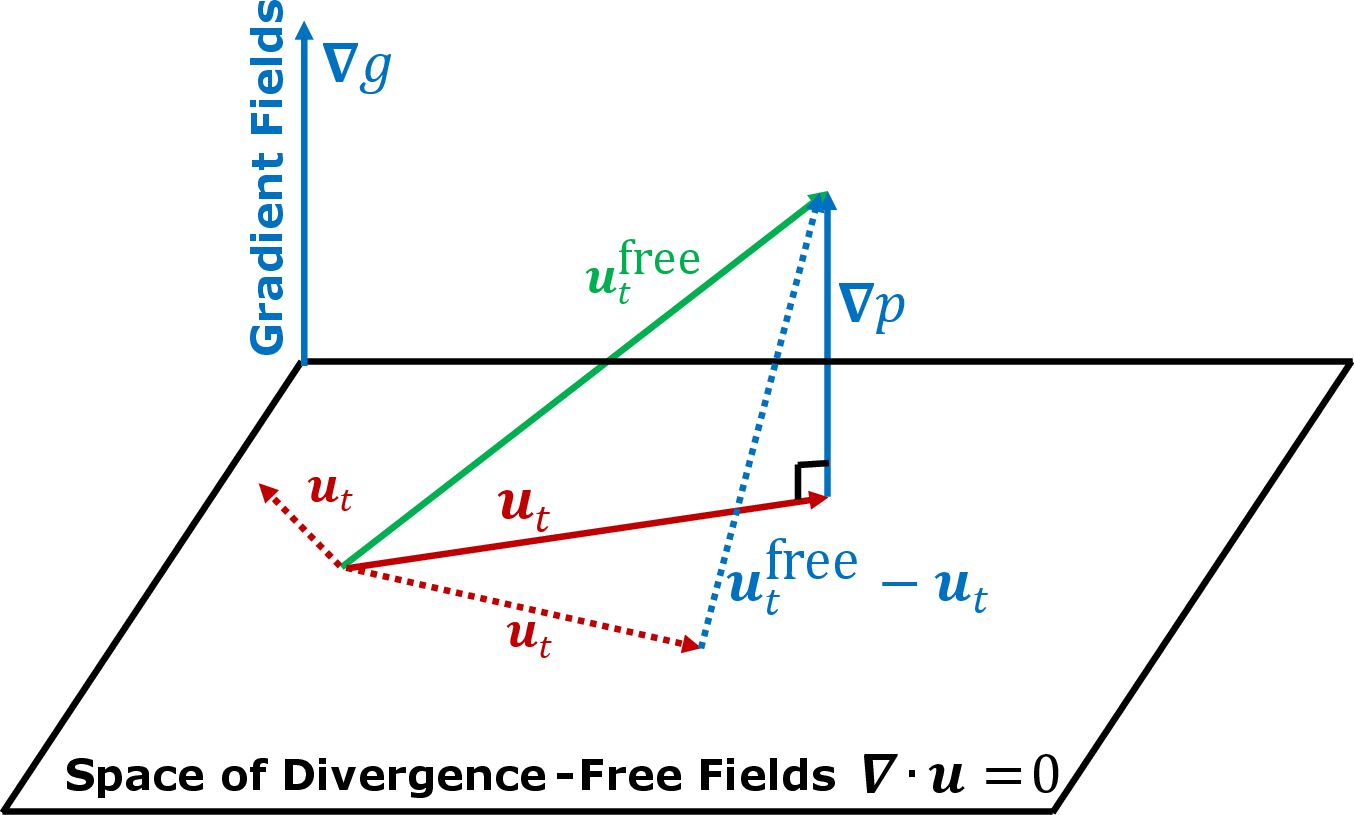}}
\end{array}$
\caption{A Schematic diagram illustrating the Helmholtz-Leray projection and geometry of incompressible flows.}
\label{Fig:Helmholtz}
\end{center}
\end{figure*}

Given a square integrable vector field $\bm{v}(\bm{x})$ over a smooth domain $\Omega\subset\mathbb{R}^n$, it can be \textit{uniquely} decomposed into two orthogonal components: (i) a divergence-free field $\bm{w}$ that satisfies the no-penetration boundary condition $\bm{w}\cdot\bm{n}=0$ on $\partial\Omega$, and (ii) a gradient field $\bm\nabla g$, for some scalar function $g$, as illustrated schematically in Fig. \ref{Fig:Helmholtz_Schematic}. Accordingly,
\[ \bm{v}(\bm{x}) = \bm{w}(\bm{x}) + \bm\nabla g (\bm{x}) \;\; \forall \; \bm{x}\in\Omega. \]
The orthogonality is understood in the $\mathbb{L}^2$ sense:
\[ \int_\Omega \left(\bm\nabla g \cdot \bm{w}\right) d\bm{x} = 0.\]
This orthogonality follows directly from integration by parts:
\begin{equation}\label{eq:Helmholtz_Orthogonality}
\int_\Omega \left(\bm\nabla g \cdot \bm{w}\right) d\bm{x} = -\int_\Omega \left( g \underbrace{\bm\nabla\cdot \bm{w}}_{=0 \; \mbox{in}\; \Omega}\right) d\bm{x} + \oint_{\partial\Omega} \left( g \underbrace{\bm{w}\cdot\bm{n}}_{=0 \; \mbox{on}\; \partial\Omega}\right) d\bm{x} = 0.
\end{equation}

This perspective is particularly illuminating for the geometry of incompressible flows. From this viewpoint, the Navier-Stokes equation may be rearranged as
\begin{equation}\label{eq:NS_Rearranged}
  \underbrace{\bm{F}-\bm{u}\cdot \bm\nabla \bm{u}}_{\bm{u}_t^{\rm{free}}} = \bm{u}_t + \frac{1}{\rho}\nabla p,
\end{equation}
Equation \eqref{eq:NS_Rearranged} reveals that the dynamic evolution of an incompressible flow from one instant to the next may be interpreted as a Helmholtz–Leray projection onto the space of divergence-free vector fields \cite{Chorin_Marsden_Book,Geometric_Control_Fluid_Dynamics,Helmholtz_Decomposition_Review}. For a given smooth velocity field $\bm{u}$ at some instant, both the convective acceleration $\bm{u}\cdot \bm\nabla \bm{u}$ and the impressed forces (e.g., viscous or gravitational) are known. Together, they define a vector field $\bm{u}_t^{\rm{free}}$ that admits a unique Helmholtz decomposition into two orthogonal components: (i) a divergence-free component satisfying the no-penetration boundary condition, represented by the local acceleration $\bm{u}_t$, and (ii) a gradient component, represented by the pressure gradient $\bm\nabla p$, as illustrated in the schematic diagram of Fig. \ref{Fig:Incompressible_Schematic}.

The schematic illustrates that, on the instantaneous subspace of admissible motions defined by the divergence-free constraint, there exist infinitely many candidates. Each candidate deviates from the \textit{free motion} $\bm{u}_t^{\rm{free}}$ and requires a constraint force to ensure incompressibility; i.e., to project $\bm{u}_t^{\rm{free}}$ onto the space of divergence-free fields. According to Gauss's principle, Nature selects the motion with the \textit{least} deviation from the free one; equivalently, it is the motion that requires the smallest possible magnitude of the constraint force required to ensure incompressibility. Geometrically, this selection corresponds to the orthogonal projection of the free motion $\bm{u}_t^{\rm{free}}$ onto the space of divergence-free fields. The associated constraint force must therefore lie in the orthogonal complement of that subspace, namely the space of gradient fields. Hence, the pressure gradient emerges as the constraint force enforcing the continuity equation and the no-penetration boundary condition. The free motion is precisely the motion that would occur in the absence of this constraint force, as given by the left-hand side of \eqref{eq:NS_Rearranged}.

On the basis of the preceding discussion, the Gaussian cost for incompressible flows (ignoring gravitational and electromagnetic forces) can be written as:
\begin{equation}\label{eq:Gauss_Incompressible}
  \mathcal{A} = \frac{1}{2} \int_\Omega \rho \left| \bm{u}_t + \bm{u}\cdot \bm\nabla \bm{u} - \nu \nabla^2 \bm{u} \right|^2 d\bm{x},
\end{equation}
where $\nu$ denotes the kinematic viscosity. This cost was proposed by Taha et al. \cite{PMPG_PoF} as the basis for the \textit{Principle of Minimum Pressure Gradient} (PMPG), which is the continuum-mechanics analogue of Gauss's principle of least constraint. It asserts that an incompressible flow evolves from one instant to the next by minimizing the magnitude of the pressure force required to ensure the continuity constraint. Any alternative evolution would require an unnecessarily larger pressure force to ensure continuity, which is against physical considerations as conceived by Gauss.

Similar to the fact that the first-order optimality condition of the Gaussian cost $Z$ in (\ref{eq:Gauss}) recovers Newton's equation of motion in particle mechanics, the first-order optimality condition of the PMPG cost functional $\mathcal{A}$ in (\ref{eq:Gauss_Incompressible}) yields the Navier-Stokes equation. In particular, any incompressible flow whose evolution minimizes $\mathcal{A}$ at every instant is guaranteed to satisfy the Navier-Stokes equation, as established in Theorem 1 in \cite{PMPG_PoF,VPNS_PRF} and Proposition 2 in \cite{NS_QP_IEEE}.

\begin{wrapfigure}{l}{0.50\textwidth}
\vspace{-0.3in}
 \begin{center}
 \includegraphics[width=7.5cm]{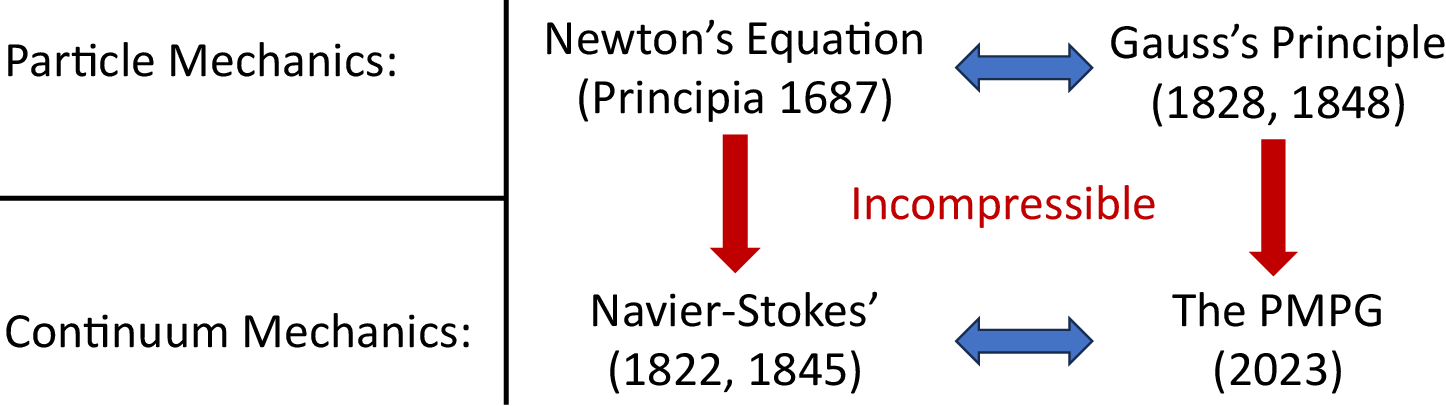}  \vspace{-0.1in}
 \caption{Schematic illustrating the equivalence between Newton's equations and Gauss's principle in particle mechanics, and the corresponding equivalence between the Navier-Stokes equation and the Principle of Minimum Pressure Gradient (PMPG) in incompressible continuum mechanics.}\vspace{-0.15in}
 \label{Fig:Gauss_Newton_PMPG}
 \end{center}\vspace{-0.2in}
\end{wrapfigure} \noindent In this sense, the Principle of Minimum Pressure Gradient plays, within incompressible continuum mechanics, a role directly analogous to that played by Gauss's principle in particle mechanics. Figure \ref{Fig:Gauss_Newton_PMPG} illustrates this correspondence by showing the equivalence between Newton's equations and Gauss's principle in particle mechanics, and the analogous relationship between the Navier-Stokes equation and the PMPG in the continuum mechanics of incompressible flows.

In the absence of all impressed forces such as viscous forces and body forces (i.e., for an ideal fluid), the PMPG, similar to Gauss's principle in particle mechanics, reduces to a purely inertial minimization. In this case, the resulting evolution corresponds to geodesic flows on the configuration manifold. This geometric interpretation is fully consistent with Arnold's seminal result: the trajectories of ideal flows are geodesics with respect to the right-invariant $\mathbb{L}^2$ (kinetic energy) Riemannian metric on the manifold of volume-preserving diffeomorphisms \cite{Arnold_French,Geometric_Control_Fluid_Dynamics_Book}.

\section{Two-Way Equivalence Between the Navier-Stokes and the Principle of Minimum Pressure Gradient}
This section contains one of the central contributions of the present paper. We extend the one-way implication established in Theorem 1 in \cite{PMPG_PoF,VPNS_PRF} and Proposition 2 in \cite{NS_QP_IEEE} to a two-way (\textit{if and only if}) equivalence. We begin by stating the assumptions that will be used throughout this section.

\textbf{General Assumptions}: Let $\rho>0$ be a constant and let $\Omega\subset\mathbb{R}^3$ be a bounded domain with smooth boundary $\partial\Omega$. Assume there exists $T>0$ such that for all $t\in[0,T]$, the velocity field $\bm{u}(\bm{x},t)$ satisfies: (i) $\bm{u}\cdot\bm\nabla \bm{u}\in\mathbb{L}^2(\Omega)$, (ii) a square integrable viscous force: $\bm\nabla \cdot \bm\tau \in \mathbb{L}^2(\Omega)$. In addition, assume a square-integrable impressed body force $\bm{f} \in \mathbb{L}^2(\Omega)$ for all $t\in[0,T]$.

\subsection{Necessary Condition for Optimality}
\textbf{Lemma 1} (cf. Theorem 1 in \cite{PMPG_PoF}): Recall the General Assumptions stated above. In addition, suppose that the initial condition $\bm{u}(\bm{x},0)$ is kinematically admissible:
\begin{equation}\label{eq:Kinematic_Admissibility}
\bm\nabla\cdot\bm{u}(\bm{x};0)=0 \; \forall \; \bm{x}\in\Omega \;\; \mbox{and}\;\; \bm{u}(\bm{x};0)\cdot\bm{n}=h(\bm{x}) \; \forall \; \bm{x}\in\partial\Omega,
\end{equation}
where $\bm{n}$ denotes the outward unit normal and $h$ is a prescribed smooth boundary function.

Suppose that for every $t\in[0,T]$, the time derivative $\bm{u}_t(\cdot,t)\in\mathbb{L}^2(\Omega)$ exists and minimizes the functional
\begin{equation}\label{eq:PMPG_Cost}
 \mathcal{A}(\bm{u}_t) = \frac{1}{2}\int_\Omega \rho \left|\bm{u}_t+\bm{u}\cdot\bm\nabla \bm{u}-\frac{1}{\rho}\left(\bm\nabla \cdot \bm\tau+\bm{f}\right)\right|^2 d\bm{x}
 \end{equation}
over the space of admissible evolutions:
\begin{equation}\label{eq:Admissible_Set}
\Theta=\{\bm{v}\in\mathbb{L}^2| \; \bm\nabla \cdot\bm{v}=0 \; \forall \; \bm{x}\in\Omega, \;\; \mbox{and} \;\; \bm{v}\cdot \bm{n} = 0\; \forall \; \bm{x}\in\partial\Omega\}.
 \end{equation}
Then $\bm{u}(\bm{x};t)$ must satisfy the Navier-Stokes equations
\begin{equation}\label{eq:Navier-Stokes}
\rho\left(\bm{u}_t+\bm{u}\cdot\bm\nabla \bm{u}\right)=-\bm\nabla \lambda+\bm\nabla \cdot \bm\tau+\bm{f} \;\; \mbox{and} \;\; \bm\nabla\cdot\bm{u}=0 \; \forall \; \bm{x}\in\Omega, \;\; t\in[0,T]
 \end{equation}
for some $\lambda\in H^1(\Omega)$, together with the no-penetration boundary condition
\begin{equation}\label{eq:No_Penetration}
 \bm{u}(\bm{x};t)\cdot\bm{n}=h(\bm{x}) \; \forall \; \bm{x}\in\partial\Omega \;\; t\in[0,T].
\end{equation}

\noindent\textbf{Proof:} Fix $t\in[0,T]$. To minimize $\mathcal{A}$ subject to the constraint $\nabla\cdot \bm{u}_t=0$ for all $\bm{x}\in\Omega$, we construct the Lagrangian
\[ \mathcal{L}(\bm{u}_t) = \mathcal{A}(\bm{u}_t) - \int_\Omega \lambda(\bm{x}) \left(\bm\nabla \cdot\bm{u}_t(\bm{x};t)\right) d\bm{x}, \]
where $\lambda$ is a Lagrange multiplier. A necessary condition for the constrained  minimization problem is then: the first variation of the Lagrangian vanishes with respect to variations in $\bm{u}_t(\bm{x})$ that belong to $\Theta$. The first variation of $\mathcal{L}$ with respect to $\bm{u}_t$ is written as
\begin{equation}\label{eq:First_Variation}
   \delta\mathcal{L} = \int_\Omega \left[\rho \left(\bm{u}_t+\bm{u}\cdot\bm\nabla \bm{u}-\frac{1}{\rho}\left(\bm\nabla \cdot \bm\tau+\bm{f}\right) \right)\cdot\delta \bm{u}_t - \lambda \bm\nabla \cdot\delta\bm{u}_t \right] d\bm{x}=0.
\end{equation}
Using the identity
\[ \lambda \bm\nabla \cdot\delta\bm{u}_t = \bm\nabla \cdot (\lambda\delta\bm{u}_t)-\bm\nabla \lambda \cdot \delta\bm{u}_t, \]
integrating the divergence term, and applying the divergence theorem we obtain
\[ \int_\Omega \bm\nabla \cdot (\lambda\delta\bm{u}_t) d\bm{x} =  \int_{\partial\Omega} \lambda\delta\bm{u}_t\cdot\bm{n} d\bm{x}. \]

Since admissible variations must satisfy $\delta\bm{u}_t\cdot \bm{n}=0$ on $\partial\Omega$, the boundary term vanishes: $\int_\Omega \bm\nabla \cdot (\lambda\delta\bm{u}_t) d\bm{x}=0$. Hence, we have
\[ \delta\mathcal{L} = \int_\Omega \left[\rho \left(\bm{u}_t+\bm{u}\cdot\bm\nabla \bm{u}\right) + \bm\nabla \lambda - \frac{1}{\rho}\left(\bm\nabla \cdot \bm\tau+\bm{f}\right) \right]\cdot \delta \bm{u}_t d\bm{x}=0 \]
Since this holds for all admissible $\delta\bm{u}_t\in\Theta$, we obtain
\[\rho\left(\bm{u}_t+\bm{u}\cdot\bm\nabla \bm{u}\right)=-\bm\nabla \lambda+\bm\nabla \cdot \bm\tau+\bm{f} \;\; \forall \; \bm{x}\in\Omega. \]
This argument holds for each $t\in[0,T]$.

In addition, because the initial condition $\bm{u}(\bm{x};0)$ is divergence-free for all $\bm{x}\in\Omega$ and the local acceleration $\bm{u}_t(\bm{x};t)$ is divergence-free for all $\bm{x}\in\Omega$ and $t\in[0,T]$, then it implies that
\[\bm\nabla\cdot \bm{u}(\bm{x},t) = 0 \;\; \forall\; \bm{x}\in\Omega \;\; \mbox{and} \; t\in[0,T].\]
Similarly, since the initial condition $\bm{u}(\bm{x};0)$ satisfies the no-penetration boundary condition
\[ \bm{u}(\bm{x};0)\cdot\bm{n}=h(\bm{x}) \;\; \forall \bm{x}\in\partial\Omega,\]
and the local acceleration $\bm{u}_t(\bm{x};t)$ satisfies a homogenous normal boundary condition
\[ \bm{u}_t(\bm{x};t)\cdot\bm{n}=0 \;\; \forall \bm{x}\in\partial\Omega \;\; \mbox{and} \; t\in[0,T],\]
then it implies that
\[ \bm{u}(\bm{x};t)\cdot\bm{n}=h(\bm{x}) \;\; \forall \bm{x}\in\partial\Omega \;\; \mbox{and} \; t\in[0,T],\]
which completes the proof. $\blacksquare$\\

\textbf{Verbal Statement of Lemma 1:} \textit{Among all kinematically-admissible flows---namely, those that satisfy the continuity constraint and the prescribed normal boundary condition (\ref{eq:Kinematic_Admissibility})---the flow whose evolution $\bm{u}_t$ minimizes the PMPG cost functional \eqref{eq:PMPG_Cost} at each instant is guaranteed to satisfy the Navier-Stokes equations}.\\

\textbf{Remark 1:} The above result is independent of spatial dimension and applies to both inviscid and viscous flows. It accommodates arbitrary viscous stress models that do not depend explicitly on the local acceleration $\bm{u}_t$, as well as arbitrary square-integrable body forces $\bm{f}$. The essential structural requirements are incompressibility and a prescribed normal boundary condition (e.g., no-penetration).\\

\textbf{Remark 2:} The above result extends to time-dependent domains $\Omega(t)\subset\mathbb{R}^3$ with sufficiently smooth boundary motion. In this setting, the no-penetration condition becomes
\begin{equation}
\bm{u}(\bm{x},t)\cdot\bm{n}(\bm{x},t)
=
\bm{V}_b(\bm{x},t)\cdot\bm{n}(\bm{x},t),
\qquad
\bm{x}\in\partial\Omega(t),
\end{equation}
where $\bm{V}_b$ denotes the prescribed boundary velocity. Differentiating this condition along the moving boundary yields a prescribed normal component of the local acceleration,
\begin{equation}
\bm{u}_t(\bm{x},t)\cdot\bm{n}(\bm{x},t)
=
\gamma(\bm{x},t),
\qquad
\bm{x}\in\partial\Omega(t),
\end{equation}
where $\gamma$ is a known function determined by the boundary motion and the current velocity field. The admissible set therefore becomes
\begin{equation}
\Theta(t)
=
\left\{
\bm{v}\in \mathbb{L}^2(\Omega(t))
\;\middle|\;
\nabla\cdot\bm{v}=0 \text{ in } \Omega(t),
\;
\bm{v}\cdot\bm{n}=\gamma \text{ on } \partial\Omega(t)
\right\}.
\end{equation}
Since the normal component of $\bm{u}_t$ is prescribed, admissible variations satisfy
\[
\delta\bm{u}_t\cdot\bm{n}=0
\quad \text{on } \partial\Omega(t),
\]
and the variational argument proceeds unchanged.\\

\textbf{Remark 3:} The above result does not require a specific tangential boundary condition (e.g., no-slip). The equivalence holds for arbitrary tangential boundary conditions compatible with the admissible function space. See Corollary 1 below.\\

\textbf{Corollary 1}: Recall the General Assumptions above, and assume that the initial condition is kinematically admissible with a no-slip boundary condition:
\begin{equation}\label{eq:Kinematic_Admissibility_No_slip}
\bm\nabla\cdot\bm{u}(\bm{x};0)=0 \; \forall \; \bm{x}\in\Omega \;\; \mbox{and}\;\; \bm{u}(\bm{x};0)=\bm{0} \; \forall \; \bm{x}\in\partial\Omega.
\end{equation}
Assume that for every $t\in[0,T]$, the local acceleration $\bm{u}_t(\cdot,t)\in \mathbb{L}^2(\Omega)$ minimizes the functional \eqref{eq:PMPG_Cost} over the admissible set
\begin{equation}\label{eq:Admissible_Set_No_Slip}
\Theta_0=\{\bm{v}\in H^1| \; \bm\nabla \cdot\bm{v}=0 \; \forall \; \bm{x}\in\Omega, \;\; \mbox{and} \;\; \bm{v} = \bm{0}\; \forall \; \bm{x}\in\partial\Omega\},
 \end{equation}
Then $\bm{u}$ necessarily satisfies the Navier--Stokes equations \eqref{eq:Navier-Stokes} for some $\lambda\in H^1(\Omega)$, together with the no-slip boundary condition
\begin{equation}\label{eq:No_Penetration_No_slip}
 \bm{u}(\bm{x};t)=\bm{0} \; \forall \; \bm{x}\in\partial\Omega \;\; t\in[0,T].
\end{equation}

\noindent\textbf{Proof.} The proof is identical to that of Lemma~1, with the admissible set $\Theta$ replaced by $\Theta_0$. In particular, admissible variations satisfy $\delta\bm{u}_t=\bm{0}$ on $\partial\Omega$, hence $\delta\bm{u}_t\cdot\bm{n}=0$ on $\partial\Omega$, and the boundary term arising from integration by parts vanishes. $\blacksquare$\\

\textbf{Remark 4:} The corollary extends verbatim to non-homogeneous no-slip boundary conditions $\bm{u}=\bm{u}_b$ on $\partial\Omega$, provided $\bm{u}_b$ is prescribed and sufficiently regular.\\

\textbf{Remark 5:} Within this formulation, the pressure arises naturally as the Lagrange multiplier $\lambda$ enforcing the kinematic constraints of incompressibility and no-penetration. This interpretation is consistent with the classical projection and variational formulations of incompressible flow \cite{Lanczos_Variational_Mechanics_Book,Chorin_Marsden_Book,Temam_Projection,Pressure_BCs,Chorin_Projection,Projection_Book,Projection_Review,Hirsch_Book2,Geometric_Control_Fluid_Dynamics,badin2018variational,Morrison2020lagrangian,DeVoria_Hamiltonian_JFM}. Consequently, in the language of analytical mechanics, the pressure force represents the constraint force required to maintain these constraints.

\subsection{Existence of Unique Minimizers}
\textbf{Lemma 2:} Fix $t\in[0,T]$ and assume the General Assumptions hold at time $t$. Then there exists a unique minimizer $\bm{u}_t^*(\cdot,t)\in\Theta$ of the functional $\mathcal{A}$ in \eqref{eq:PMPG_Cost} over the admissible set $\Theta$ defined in \eqref{eq:Admissible_Set}.

\noindent\textbf{Proof:} Fix $t\in[0,T]$ and define the known vector field
\begin{equation}\label{eq:Free_Acceleration}
\bm{u}_t^{\rm{free}}(t):=
-\bm{u}\cdot\nabla\bm{u}
+\frac{1}{\rho}\big(\nabla\cdot\bm{\tau}+\bm{f}\big) \; \in \mathbb{L}^2(\Omega).
\end{equation}
Then the functional \eqref{eq:PMPG_Cost} can be written as
\[
\mathcal{A}(\bm{v})
=\frac{1}{2}\rho\int_\Omega |\bm{v}-\bm{u}_t^{\rm{free}}(t)|^2\,d\bm{x}
=\frac{1}{2}\rho\|\bm{v}-\bm{u}_t^{\rm{free}}(t)\|_{\mathbb{L}^2(\Omega)}^2.
\]
Hence, minimizing $\mathcal{A}$ over $\Theta$ is equivalent to finding the element of $\Theta$
closest to $\bm{u}_t^{\rm{free}}(t)$ in the $\mathbb{L}^2(\Omega)$ norm.

If $\Theta\subset \mathbb{L}^2(\Omega)$ is a closed subspace, the result follows directly from the Projection Theorem in Hilbert spaces (Theorem II.3 in \cite{Functional_Analysis}): if $\mathcal{H}$ is a Hilbert space and $\mathcal{M}\subset\mathcal{H}$ is a closed subspace, then for every $x\in\mathcal{H}$ there
exists a unique $m\in\mathcal{M}$ minimizing $J(m)=\|m-x\|_{\mathcal{H}}^2$ over $\mathcal{M}$.

$\Theta$ is a linear subspace of $\mathbb{L}^2(\Omega)$: for $\bm{v}_1,\bm{v}_2\in\Theta$ and $\alpha,\beta\in\mathbb{R}$, we have $\alpha\bm{v}_1+\beta\bm{v}_2\in\Theta$. It remains to show below that $\Theta$ is closed in $\mathbb{L}^2(\Omega)$. Let $\{\bm{v}_k\}\subset\Theta$ converge to $\bm{v}$ in $\mathbb{L}^2(\Omega)$. We show that $\bm{v}\in\Theta$.

First, for any $\psi\in C_0^\infty(\Omega)$, convergence in $\mathbb{L}^2(\Omega)$ implies
\[
\int_\Omega \bm{v}\cdot\nabla\psi\,d\bm{x}
=
\lim_{k\to\infty}\int_\Omega \bm{v}_k\cdot\nabla\psi\,d\bm{x}
=0,
\]
since each $\bm{v}_k\in\Theta$ is divergence-free in the weak sense. Hence $\nabla\cdot\bm{v}=0$ in the
distributional sense, and therefore $\nabla\cdot\bm{v}=0\in \mathbb{L}^2(\Omega)$. In particular, $\bm{v}\in H(\mathrm{div};\Omega)$, where
\[ H(\mathrm{div};\Omega)=\{\bm{w}\in \mathbb{L}^2(\Omega) |\; \bm\nabla \cdot\bm{w} \in \mathbb{L}^2(\Omega)\}. \]

Moreover, since $\nabla\cdot\bm{v}_k=0$ for all $k$,
\[
\|\bm{v}_k-\bm{v}\|_{H(\mathrm{div};\Omega)}^2
=
\|\bm{v}_k-\bm{v}\|_{\mathbb{L}^2(\Omega)}^2
+
\|\nabla\cdot(\bm{v}_k-\bm{v})\|_{\mathbb{L}^2(\Omega)}^2
=
\|\bm{v}_k-\bm{v}\|_{\mathbb{L}^2(\Omega)}^2 \to 0,
\]
so $\bm{v}_k\to\bm{v}$ in $H(\mathrm{div};\Omega)$. Since the normal trace map is continuous
from $H(\mathrm{div};\Omega)$ to $H^{-1/2}(\partial\Omega)$ \cite{Finite_Element_NS,Finite_Element_Maxwell},
it follows that
\[
\bm{v}_k\cdot\bm{n}\to \bm{v}\cdot\bm{n}
\quad \text{in } H^{-1/2}(\partial\Omega).
\]
Because $\bm{v}_k\cdot\bm{n}=0$ on $\partial\Omega$ for all $k$, we conclude $\bm{v}\cdot\bm{n}=0$ as well.
Therefore $\bm{v}\in\Theta$, and $\Theta$ is closed in $\mathbb{L}^2(\Omega)$.

The projection theorem now implies that there exists a unique minimizer $\bm{u}_t^*\in\Theta$ of
$\mathcal{A}$ at each $t\in[0,T]$. \hfill$\blacksquare$\\

\textbf{Verbal Statement of Lemma 2:} \textit{At any given instant where the $\mathbb{L}^2$-norms of the convective and viscous accelerations as well as the body force are well-defined, there exists a unique evolution $\bm{u}_t$ that minimizes the PMPG cost functional \eqref{eq:PMPG_Cost} over all kinematically-admissible evolutions}.\\

\textbf{Remark 6:} Lemma 2 is, in essence, a statement about the Helmholtz-Leray projection (e.g., \cite{Functional_Analysis_NS}) of the free acceleration $\bm{u}_t^{\rm{free}}$, given in Eq. (\ref{eq:Free_Acceleration}), onto the space of divergence-free, no-penetration fields $\Theta$. The minimizer $\bm{u}_t^*$ is precisely the $L^2$-orthogonal projection of this free acceleration. Furthermore, the Sobolev regularity and smoothness of the minimizer $\bm{u}_t^*$ matches those of the free acceleration since the Helmholtz-Leray projection is continuous in the corresponding Sobolev spaces (see, e.g., \cite{Temam_Book}).\\

\textbf{Remark 7:} Lemma 2 extends to the no-slip admissible set $\Theta_0$ defined in \eqref{eq:Admissible_Set_No_Slip}. In this case, one may define
\[
V
=
\left\{
\bm{w}\in C_0^\infty(\Omega)^3
\;\middle|\;
\nabla\cdot \bm{w}=0
\right\}.
\]
It is well known that $V$ is dense in the space of solenoidal $\mathbb{L}^2(\Omega)$ fields satisfying the homogeneous Dirichlet boundary condition (see, e.g., \cite{Finite_Element_NS}). Consequently,
\[
\Theta_0 = \overline{V}^{\,L^2(\Omega)}
\]
is a closed subspace of $\mathbb{L}^2(\Omega)$, and the projection theorem applies verbatim to yield existence and uniqueness of the minimizer within $\Theta_0$.

\subsection{Local Existence and Uniqueness of Smooth Solutions of the Navier-Stokes Equation}
We recall the following classical local well-posedness result for the three-dimensional Navier--Stokes equations (see, e.g.,  Theorem 6.2 in \cite{Galdi_Navier_Stokes,Heywood_Navier_Stokes}).\\

\textbf{Theorem 1 (Local existence and uniqueness \cite{Galdi_Navier_Stokes}):} Let $\Omega\subset\mathbb{R}^{3}$ be a domain with a smooth boundary $\partial\Omega$ (of class $C^{\infty}$). Let $\bm{u}_0$ be a smooth initial condition satisfying the compatibility condition \eqref{eq:Kinematic_Admissibility_No_slip}. Then, there exists a time $T>0$ and a \textbf{unique} pair $(\bm{u},p)$ of smooth functions that satisfy the Navier-Stokes equations (\ref{eq:Navier-Stokes}) with a Newtonian viscous stress $\bm\nabla \cdot\bm\tau=\rho\nu\bm\nabla^2 \bm{u}$ for given $\rho>0$, $\nu>0$ and $\bm{f}\equiv0$, as well as the boundary condition (\ref{eq:No_Penetration_No_slip}) over the time interval $t\in[0,T]$, matching the initial data: $\bm{u}(\bm{x};0)=\bm{u}_0$.

\noindent\textbf{Proof:} See Galdi \cite{Galdi_Navier_Stokes} and Heywood \cite{Heywood_Navier_Stokes}.\\

\textbf{Verbal Statement of Theorem 1:} \textit{Given a smooth divergence-free initial condition satisfying the no-slip boundary condition, the three-dimensional Navier--Stokes equations admit a unique smooth solution on a short time interval $[0,T]$.}\\

\textbf{Remark 8:} The local existence and uniqueness result of Theorem 1 extends to sufficiently smooth non-homogeneous Dirichlet boundary conditions, provided that the boundary data satisfy the usual compatibility conditions with the initial data (see, e.g., \cite{Galdi_Navier_Stokes}).

\subsection{Main Theorem: Two-Way Equivalence Between the PMPG and the Navier-Stokes Equation}
\textbf{Theorem 2:} Let $\bm{u}(\bm{x};t)$ be a smooth ($C^\infty$) flow field over a bounded domain $\Omega\subset\mathbb{R}^3$ and a time interval $[0,T]$. Then $\bm{u}$ is a solution of the incompressible Navier-Stokes system
\[ \begin{array}{lll}
\rho\left(\bm{u}_t+\bm{u}\cdot\bm\nabla \bm{u}\right)&=&-\bm\nabla p+ \rho\nu\bm\nabla^2 \bm{u}, \; \forall \; \bm{x}\in\Omega, \;\; t\in[0,T]\\
\bm\nabla\cdot\bm{u}&=&0, \; \forall \; \bm{x}\in\Omega, \;\; t\in[0,T] \\
\bm{u}&=&\bm{0},\; \forall \; \bm{x}\in\partial\Omega, \;\; t\in[0,T]
\end{array} \]
\textbf{\textit{if and only if}}, at every instant $t\in[0,T]$, its evolution $\bm{u}_t(\cdot;t)$ minimizes the functional
\begin{equation}\label{eq:PMPG_Cost_Newtonian}
\mathcal{A}(\bm{v}) = \frac{1}{2}\int_\Omega \rho \left|\bm{v}+\bm{u}\cdot\bm\nabla \bm{u}-\nu\bm\nabla^2\bm{u}\right|^2 d\bm{x}
\end{equation}
over the space of admissible solutions $\Theta_0$, defined in Eq. (\ref{eq:Admissible_Set_No_Slip}).

\noindent\textbf{Proof:} The sufficiency (minimization implies Navier-Stokes) follows from Corollary 1.

The converse statement follows from Lemma 2 and Theorem 1, as follows. Let $\bm{u}^{(s)}$ be a smooth solution of the Navier-Stokes with the prescribed initial and boundary data, whose existence is guaranteed by Theorem 1. Since $\bm{u}^{(s)}$ is smooth on a bounded domain, the General Assumptions are satisfied for all $t\in[0,T]$. Then, for each fixed $t\in[0,T]$, Lemma 2 (together with Remark 7) guarantees the existence of a unique minimizer
\[
\bm{u}_t^*(\cdot,t)
=
\underset{\Theta_0}{\operatorname{argmin}}
\;\mathcal{A}(\bm{v}(\cdot,t)),
\]
which is the Helmholtz-projection of the free acceleration
\[ \bm{u}_t^{\rm{free}}=-\bm{u}\cdot\bm\nabla \bm{u}+\nu\bm\nabla^2\bm{u}.\]
Since $\bm{u}\in C^\infty\Rightarrow\bm{u}_t^{\rm{free}}\in C^\infty$, then the resulting minimizer $\bm{u}_t^*$ is also smooth.

Starting from the same initial data $\bm{u}^{(s)}(\bm{x},0)$, the optimal evolution $\bm{u}_t^*\in C^\infty$ gives rise to a smooth velocity field
\[ \bm{u}^*(\bm{x},t) = \bm{u}^{(s)}(\bm{x},0) + \int_0^t \bm{u}_t^* (\bm{x},\tau) d\tau.\]
Moreover, this solution is guaranteed to satisfy the Navier-Stokes equation and boundary conditions by Lemma 1. But Theorem 1 ensures uniqueness of Navier-Stokes' solutions with the same initial and boundary data. Hence, $\bm{u}^{(s)}\equiv\bm{u}^*$, which concludes the proof. $\blacksquare$

\textbf{Verbal Statement of Theorem 2:} \textit{A candidate smooth flow field over a smooth bounded domain $\Omega\subset\mathbb{R}^3$ is a solution of the incompressible Navier-Stokes system on the time interval $[0,T]$ \textbf{if and only if}, at every instant of time, its evolution $\bm{u}_t$ minimizes the PMPG cost functional over all kinematically-admissible evolutions}.

\section{Physical Implications of the Two-Way Equivalence}
Theorem 2 establishes a strict two-way equivalence (\textit{if and only if}) between the Principle of Minimum Pressure Gradient and the Navier-Stokes equation within the class of smooth incompressible flows. An incompressible flow whose local acceleration $\bm{u}_t$ minimizes the PMPG cost functional at every instant is guaranteed to satisfy the Navier-Stokes system. Conversely, if at some instant the local acceleration of a smooth flow field requires a larger pressure force than another kinematically-admissible evolution, then that flow cannot be a solution of the Navier-Stokes equation.

This two-way equivalence can be used to shed light on the causal mechanisms behind incompressible-flow phenomena (e.g., separation, transition, etc). While the Navier--Stokes formulation and the PMPG formulation are mathematically equivalent and therefore yield identical flow fields for the same smooth initial and boundary data, the minimization framework can provide further insight into the flow behavior. In the classical formulation, the governing equations describe a balance of forces. In the PMPG formulation, the same dynamics can be interpreted as the selection, at each instant, of the admissible evolution requiring the smallest $\mathbb{L}^2$-magnitude of the pressure force. In this sense, if a flow exhibits a particular instantaneous behavior, that behavior can be viewed as the result of minimizing the constraint force among all kinematically-admissible alternatives. Such an interpretation parallels familiar reasoning in design optimization: when the optimization algorithm converges to a specific \textit{design}, it is selected because any admissible alternative would incur a strictly larger value of the cost functional. The equivalence theorem ensures that this interpretation is not heuristic, but a mathematically exact restatement of the Navier-Stokes dynamics.

For example, consider a smooth incompressible flow field $\bm{u}(\bm{x},t)$ over a curved surface at some instant $t$. Suppose that the Navier--Stokes acceleration $\bm{u}_t$ produces, at the subsequent instant $t+\Delta t$, a configuration $\bm{u}(t+\Delta t)$ that begins to separate from some location $\bm{x}_s^*$ on the surface. The two-way equivalence then implies that the obtained acceleration $\bm{u}_t$, which dictates the subsequent separating behavior, is realized because it is the evolution that requires the smallest magnitude of the pressure force to maintain incompressibility. Any alternative admissible evolution---one that would maintain an attached flow or another that would induce separation from a different location---would demand a pressure force with a strictly larger magnitude to enforce the same constraints, and therefore would not satisfy the Navier--Stokes equation at that instant. In this sense, the separating motion is generated by the minimal-pressure evolution compatible with incompressibility. The same reasoning applies to other incompressible-flow phenomena (e.g., vorticity generation, shedding, transition, etc).

\begin{wrapfigure}{l}{0.50\textwidth}
  \begin{center}
  \includegraphics[width=6cm]{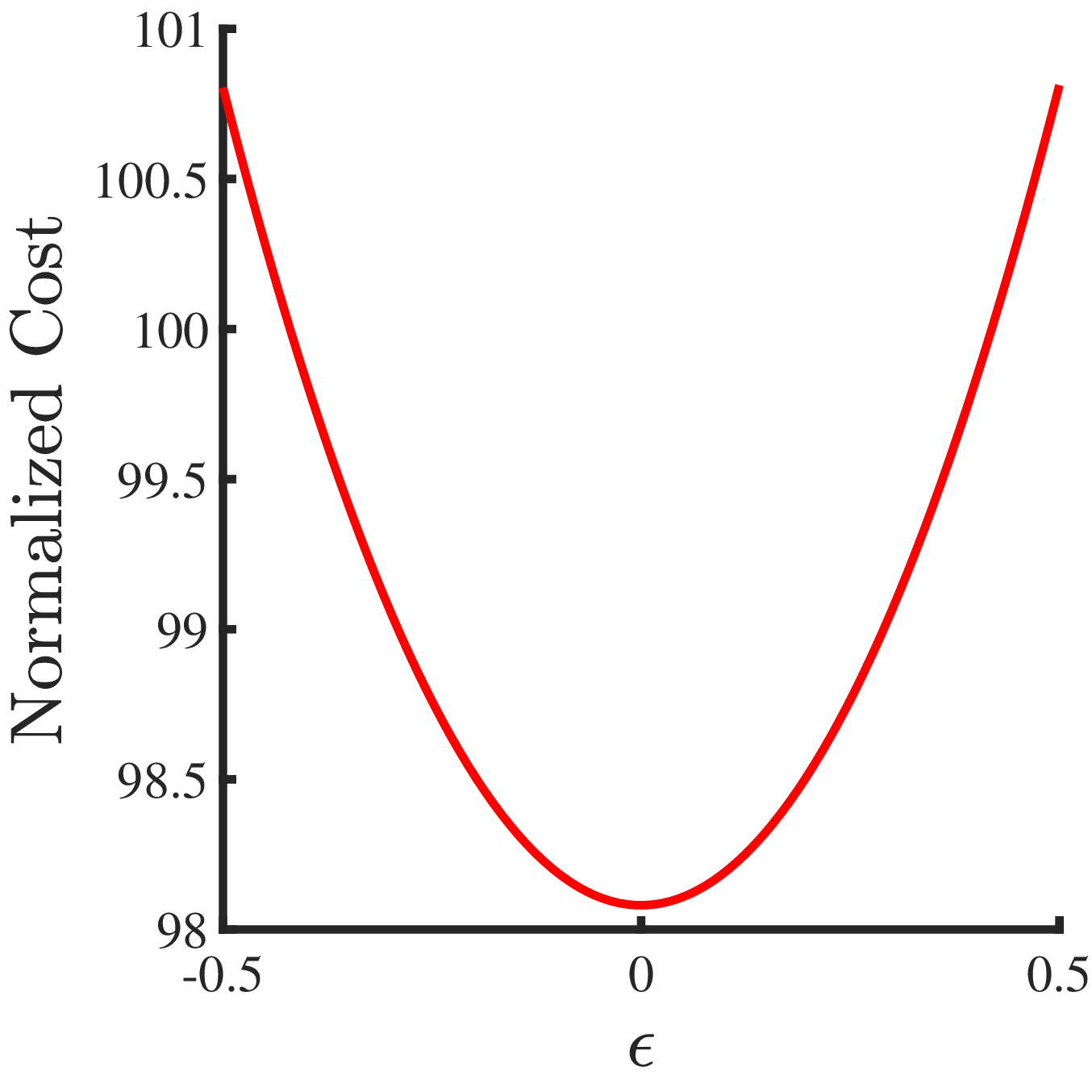}
  \caption{Variation of the normalized cost $\hat{\mathcal{A}}=\frac{\mathcal{A}}{\rho U_{lid}^4}$  with the size $\epsilon$ of perturbation from the true evolution  $\bm{u}_t^*$. The pressure gradient cost attaints its minimum precisely at $\epsilon=0$. That is, the evolution $\bm{u}_t^*$, obtained from the numerical simulation, minimizes the cost $\mathcal{A}$ over all kinematically admissible evolutions $\bm{u}_t = \bm{u}_t^* + \epsilon \bm\eta$.}\vspace{-0.1in}
  \label{Fig:Minimality_Demo1D}
  \end{center}
 \end{wrapfigure} \noindent To illustrate this perspective, we revisit the unsteady lid-driven cavity simulation presented in \cite{NS_QP_IEEE}. We select a representative instant during the simulation. At this time, the velocity field $\bm{u}$ and the corresponding evolution $\bm{u}_t^*$ (obtained from the Navier--Stokes dynamics) are available from the numerical solution. Let $\bm{\eta}$ be any kinematically-admissible perturbation; i.e., divergence-free: $\bm\nabla \cdot \bm\eta=0$ in $\Omega$, and vanishes at the boundaries of the cavity: $\bm{\eta}|_{\partial\Omega}=\bm{0}$. We then construct a family of legitimate (kinematically admissible) evolutions:
\[ \bm{u}_t = \bm{u}_t^* + \epsilon \bm\eta, \]
where $\epsilon$ is the size of perturbation from the true evolution $\bm{u}_t^*$ in the direction of $\bm\eta$. Since each member of this family satisfies the incompressibility constraint and boundary conditions, they represent legitimate alternative evolutions.

The PMPG cost functional $\mathcal{A}$ at this instant depends on the current velocity field $\bm{u}$ and the candidate evolution $\bm{u}_t$, as defined in Eq.~(\ref{eq:Gauss_Incompressible}). We then compute $\mathcal{A}(\bm{u}_t;\bm{u})$ for each member of this family. Figure~\ref{Fig:Minimality_Demo1D} shows the variation of the normalized PMPG cost with $\epsilon$. The functional attains its minimum precisely at $\epsilon=0$, corresponding to the Navier--Stokes evolution $\bm{u}_t^*$, while any nonzero perturbation produces a strictly larger cost.

\begin{figure*}
\begin{center}
$\begin{array}{cc}
\subfigure[Nondimensional $\bm\eta_1$.]{\label{fig:var_Udot_per1} \includegraphics[width=6.5cm]{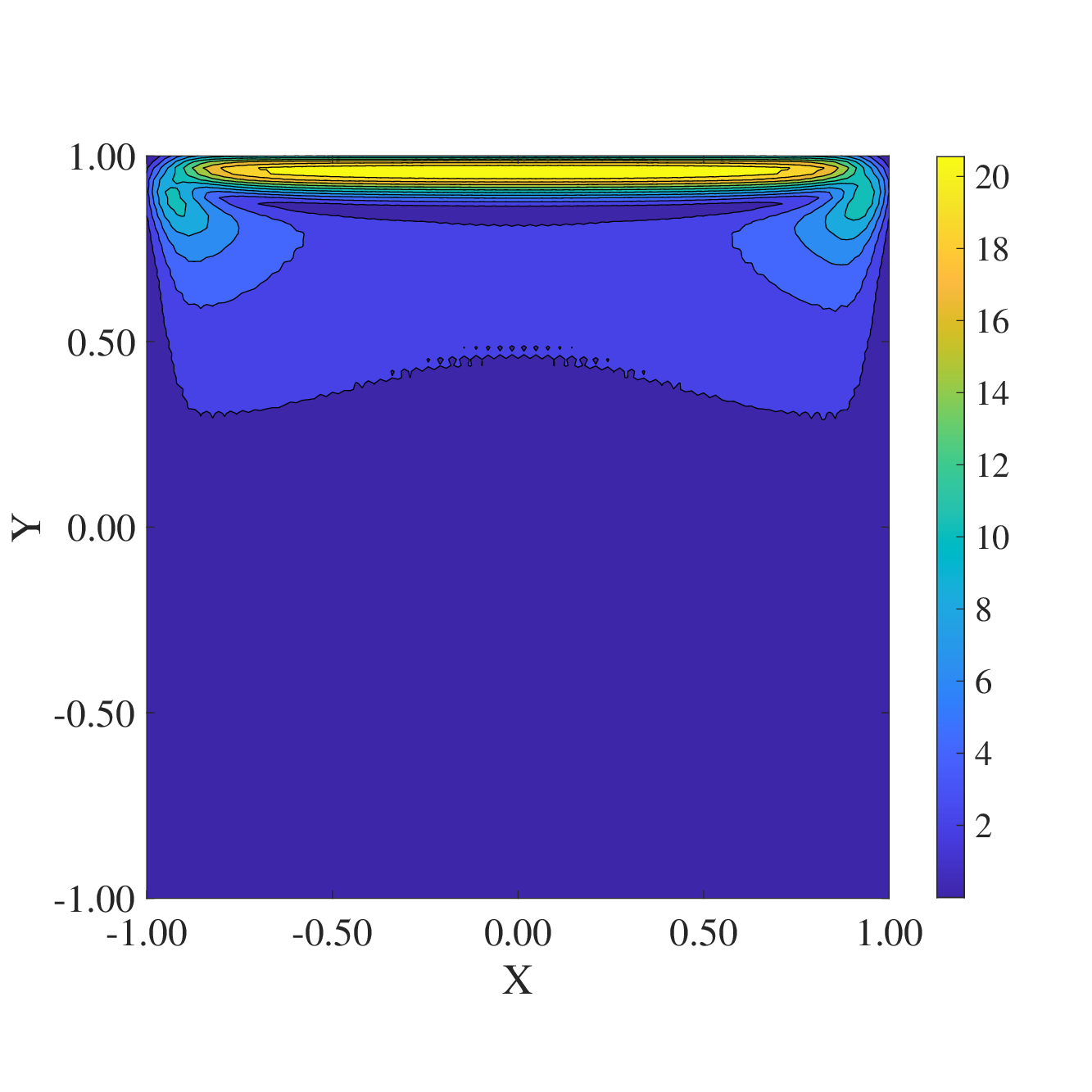}}
& \subfigure[Nondimensional $\bm\eta_2$.]{\label{fig:var_Udot_per2}\includegraphics[width=6.5cm]{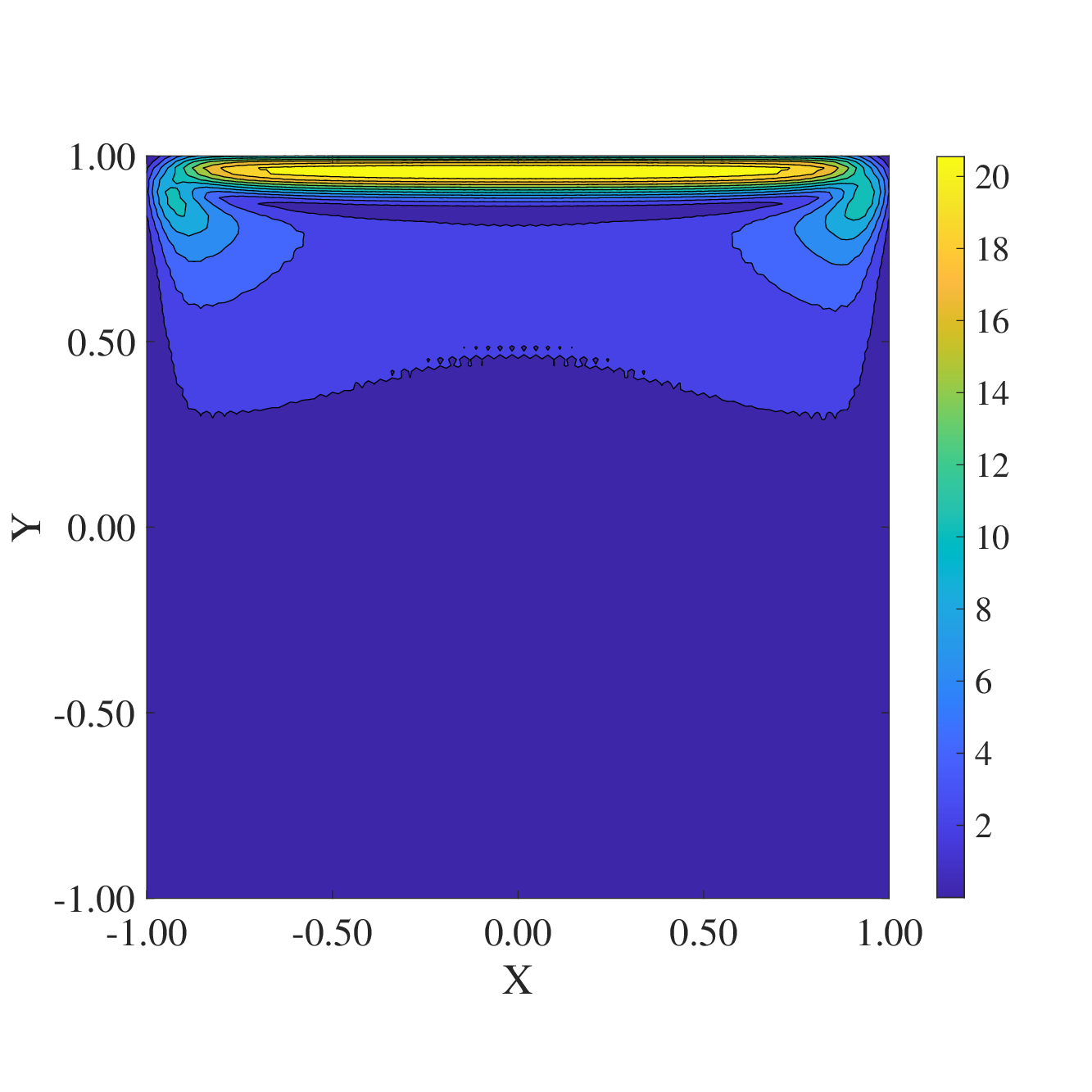}}\\

\subfigure[Nondimensional $\dot{\bm{U}}^*$.]{\label{fig:var_Udot_tby2} \includegraphics[width=6.5cm]{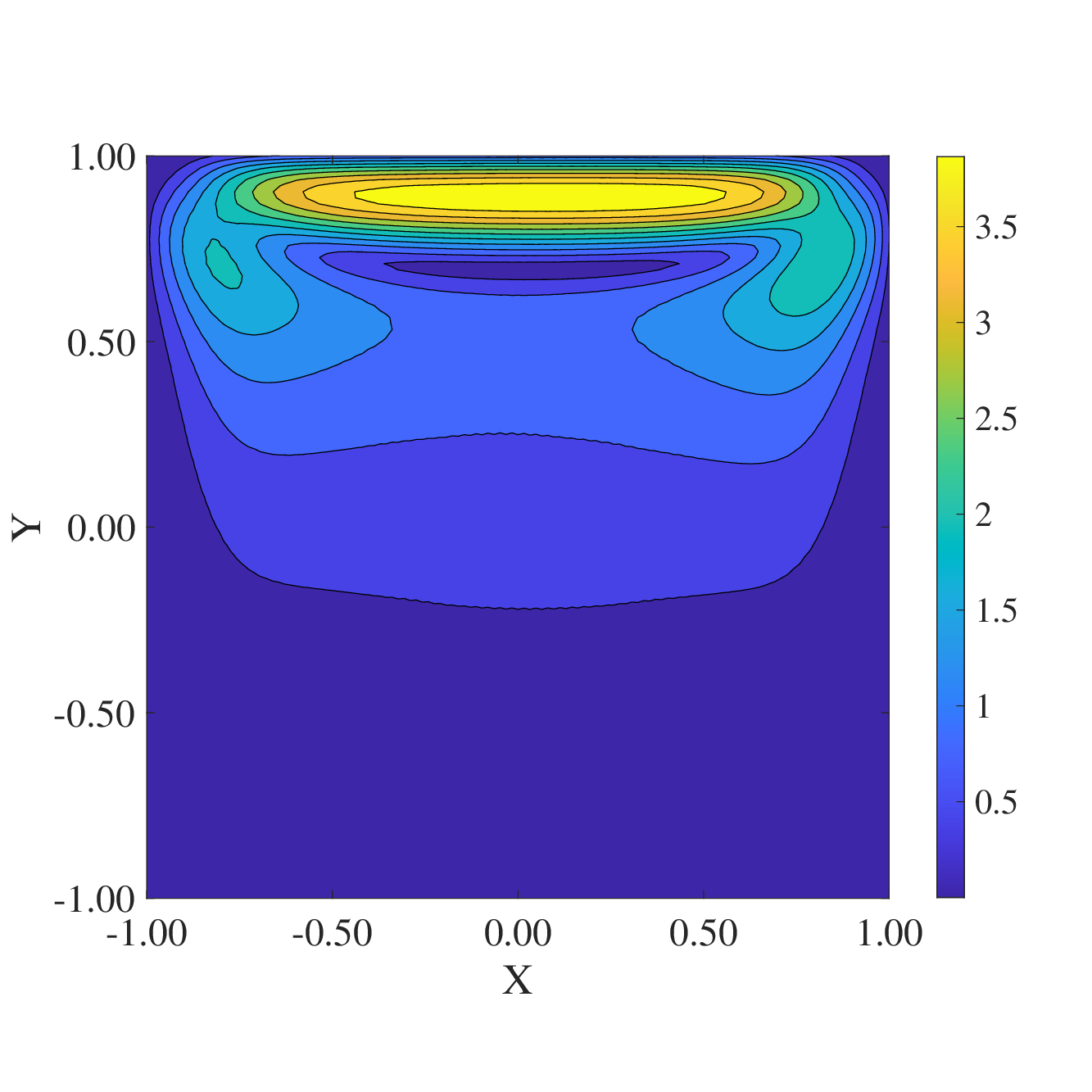}}
& \subfigure[Normalized Cost $\hat{\mathcal{A}}$.]{\label{fig:App2DPert}\includegraphics[width=6.5cm]{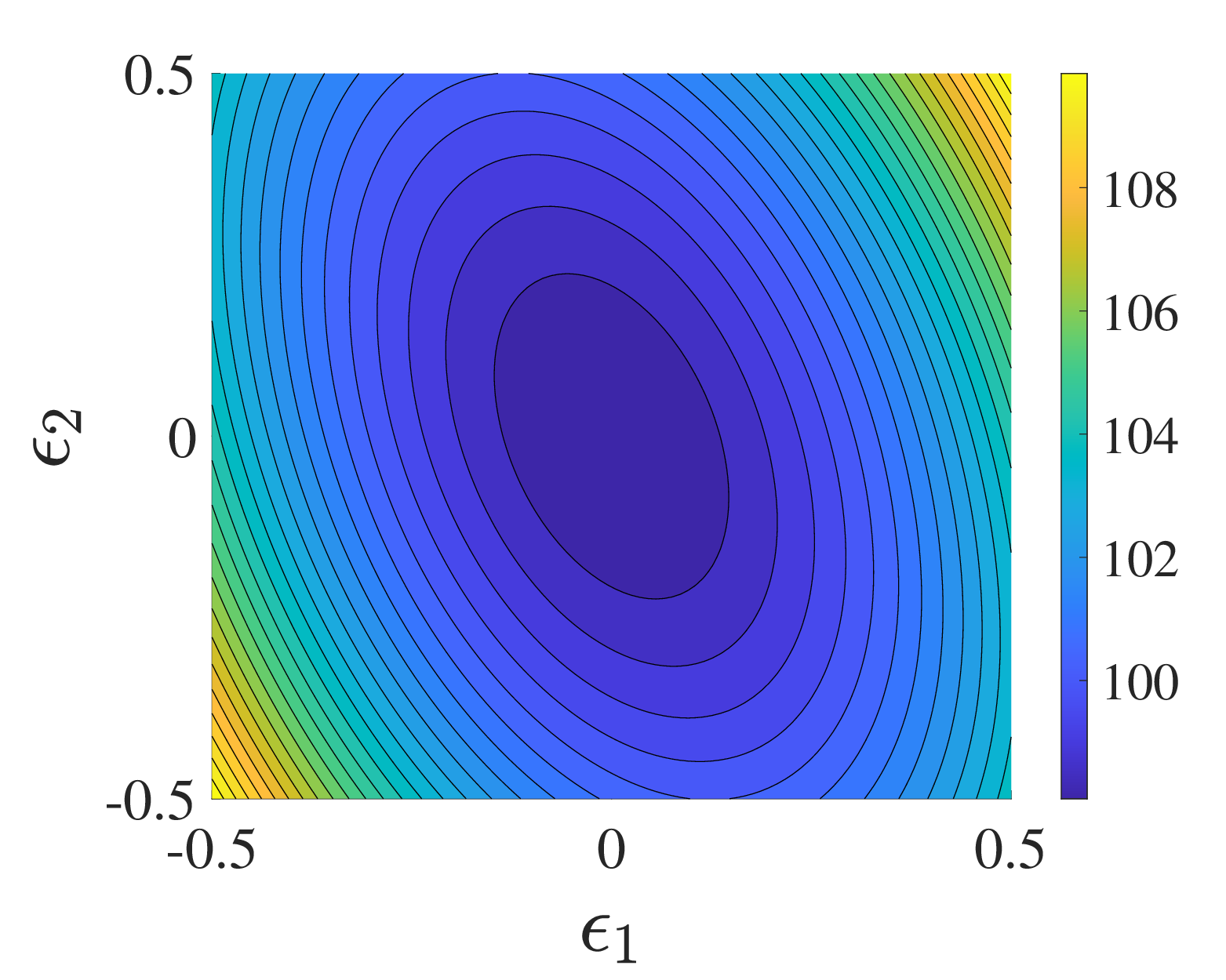}}

\end{array}$
\caption{Contours of the magnitude of the nondimensional evolution $\bm{u}_t \frac{T_{ref}}{U_{lid}}$ corresponding to (a, b) the perturbations $\bm\eta_1$, $\bm\eta_2$, and (c) the true evolution $\bm{u}_t^*$. The subfigure (d) shows contours of the nondimensional cost $\hat{\mathcal{A}}$ in the $\epsilon_1$-$\epsilon_2$ plane. The PMPG cost attains its minimum precisely at $\epsilon_1=\epsilon_2=0$, confirming the optimality of $\bm{u}_t^*$ over the family $\bm{u}_t = \bm{u}_t^* + \epsilon_1 \bm\eta_1 + \epsilon_2 \bm\eta_2$.}
\label{Fig:Minmiality_Demo2D}
\end{center}
\end{figure*}

A similar behavior is obtained for other perturbation directions. For example, let $\bm{\eta}_1$ and $\bm{\eta}_2$ be two kinematically-admissible perturbations satisfying $\nabla\cdot\bm{\eta}_i=0$ in $\Omega$ and $\bm{\eta}_i|_{\partial\Omega}=\bm{0}$ for $i=1,2$.
We construct the two-parameter family of admissible instantaneous evolutions:
\[ \bm{u}_t = \bm{u}_t^* + \epsilon_1 \bm\eta_1 + \epsilon_2 \bm\eta_2,\]
where $\epsilon_1$ and $\epsilon_2$ control the perturbation magnitudes. We then evaluate the PMPG cost functional $\mathcal{A}(\bm{u}_t;\bm{u})$ for each member of this family. Figure \ref{Fig:Minmiality_Demo2D} shows contours of $\bm{u}_t^*$, $\bm\eta_1$, and $\bm\eta_2$ in addition to the contours of the normalized cost in the plane $\epsilon_1$-$\epsilon_2$. The PMPG cost attains its minimum precisely at $\epsilon_1=\epsilon_2=0$,  confirming that $\bm{u}_t^*$ is the unique minimizer within this perturbation subspace.

This perspective also suggests a possible diagnostic criterion for assessing numerical solvers for incompressible flow simulation. Suppose two solvers enforce incompressibility to comparable tolerance levels and are applied to the same discretization and boundary-value problem. If one solver consistently produces a smaller PMPG cost $\mathcal{A}$ than the other, then it may be regarded as more faithfully replicating the underlying flow dynamics. In this sense, the PMPG cost provides not only a theoretical principle but also a quantitative measure of dynamical consistency.

\section{Finite-Dimensional Approximation and Connection to Galerkin Projection}\label{Sec:Galerkin}
It is instructive to observe that applying the PMPG in a finite-dimensional modal setting, with divergence-free modes, is equivalent to classical Galerkin projection. Consider the standard modal expansion of the velocity field
\begin{equation}\label{eq:Modal_Representation}
\bm{u}(\bm{x},t) = \sum_{i=1}^n \alpha_i(t) \bm\psi_i (\bm{x}),
\end{equation}
where the modes $\bm\psi_i$ are divergence free:
\[ \bm\nabla \cdot \bm\psi_i (\bm{x})= 0 \;\; \forall\; \bm{x}\in\Omega, \; i\in\{1, ..., n\}. \]
The PMPG formulation then reduces to the finite-dimensional optimization problem:
\begin{equation}\label{eq:PMPG_Cost_Modal}
\min_{\dot\alpha_i} \;\; \mathcal{A} (\dot\alpha_i) = \frac{1}{2}\int_\Omega \rho \left|\dot\alpha_i \bm\psi_i + \alpha_j\alpha_k\bm\psi_j\cdot\bm\nabla \bm\psi_k-\nu\alpha_j\bm\nabla^2 \bm\psi_j\right|^2 d\bm{x},
 \end{equation}
where Einstein summation is used.

This minimization problem is unconstrained (assuming that the modes satisfy the geometric boundary conditions), because the representation (\ref{eq:Modal_Representation}) automatically enforces incompressibility. The optimization is therefore taken directly with respect to the coefficients $\dot{\alpha}_i$, and the first-order necessary condition for optimality is:
\[ \frac{\partial \mathcal{A} }{ \partial \alpha_i } = 0 \; \Rightarrow \;  \int_\Omega \bm\psi_i \cdot \left(\dot\alpha_j \bm\psi_j + \alpha_j\alpha_k\bm\psi_j\cdot\bm\nabla \bm\psi_k-\nu\alpha_j\bm\nabla^2 \bm\psi_j\right) d\bm{x} = 0,\]
which leads to the finite-dimensional system of ordinary differential equations:
\begin{equation}\label{eq:Modal_ODE_Dynamics}
M_{ij} \dot\alpha_j + C_{ijk} \alpha_j \alpha_k - \nu L_{ij} \alpha_j = 0,
\end{equation}
where
\[ M_{ij} = \int_\Omega \bm\psi_i \cdot \bm\psi_j d\bm{x}, \; C_{ijk}=\int_\Omega \bm\psi_i \cdot\left(\bm\psi_j\cdot\bm\nabla \bm\psi_k\right) d\bm{x}, \; L_{ij} = \int_\Omega \bm\psi_i \cdot \bm\nabla^2 \bm\psi_j d\bm{x}.\]
The ODE system (\ref{eq:Modal_ODE_Dynamics}) is precisely the standard Galerkin projection of the Navier--Stokes equation onto the function space spanned by the modes $\bm\psi_i$ (e.g., \cite{Turbulence_Book_Holmes}).

If the modes, however, are not divergence-free, the two formulations (PMPG and classical Galerkin) may not lead to identical dynamical systems. In this case, the standard mixed Galerkin method introduces a pressure expansion
\[ p(\bm{x},t) = \sum_{\ell=1}^m \beta_\ell(t) q_\ell (\bm{x}),\]
where $\{q_\ell\}$ span a suitable pressure space. Incompressibility is then enforced in the weak sense by requiring
\[ \int_\Omega q_\ell \left(\bm\nabla\cdot \bm{u}\right) d\bm{x}=0 \; \forall \; \ell\in\{1,...,m\},  \]
which yields
\[ B_{\ell j} \alpha_j := \left[\int_\Omega q_\ell \left(\bm\nabla\cdot \bm\psi_j\right) d\bm{x}\right]\alpha_j=0. \]
The mixed Galerkin formulation of the momentum equation then gives the coupled system
\begin{equation}\label{eq:Galerkin_NonDivFree}
M_{ij} \dot\alpha_j + C_{ijk} \alpha_j \alpha_k = - G_{i\ell} b_\ell+ \nu L_{ij} \alpha_j , \\
\end{equation}
together with the discrete incompressibility constraint $B_{\ell j} \alpha_j=0$, where
\[ G_{i\ell} = \int_\Omega \bm\psi_i \cdot \bm\nabla q_\ell d\bm{x}.\]

In contrast, in the PMPG framework, there is no \emph{a priori} pressure expansion; rather, the pressure arises as the Lagrange multiplier enforcing incompressibility in the instantaneous minimization problem. To impose such a constraint in a finite-dimensional setting, introduce $m$ linear functionals $\{\mathcal{E}_\ell\}_{\ell=1}^m$ acting on scalar fields over $\Omega$ . We enforce the discrete continuity constraints as
\[
\mathcal{E}_\ell\!\left(\nabla\cdot \bm{u}_t\right)=0,
\qquad \ell=1,\dots,m.
\]
Using $\bm{u}_t=\dot{\alpha}_j \bm{\psi}_j$, these constraints become
\begin{equation}\label{eq:Continuity_Modal_PMPG}
D_{\ell j}\dot{\alpha}_j=0,
\qquad
D_{\ell j}:=\mathcal{E}_\ell\!\left(\nabla\cdot\bm{\psi}_j\right),
\end{equation}
or, in matrix form, $[\bm{D}]\dot{\bm{\alpha}}=0$.

However, for the constrained minimization problem to be well-posed, the admissible set must be non-empty; equivalently, the null space of $\bm{D}$ must be non-trivial. This requires that the span of the chosen velocity modes contain discrete divergence-free combinations, i.e., there exist $\dot{\bm{\alpha}}\neq \bm{0}$ such that $[\bm{D}]\dot{\bm{\alpha}}=\bm{0}$. Under this natural compatibility condition, the strictly convex quadratic functional $\mathcal{A}$ admits a unique minimizer within the constrained subspace.

The PMPG minimization problem (\ref{eq:PMPG_Cost_Modal}) is therefore carried out subject to the discrete continuity constraint (\ref{eq:Continuity_Modal_PMPG}). Over the set of admissible evolutions parameterized by $\dot\alpha_j$ that satisfy $[\bm{D}]\dot{\bm{\alpha}}=0$, the PMPG selects the unique evolution that minimizes the magnitude of the pressure force required to enforce the discrete continuity constraints. To impose these $m$ constraints, introduce $m$ Lagrange multipliers $\lambda_\ell$, and the constrained Lagrangian is written as
\[ \mathcal{L}(\dot\alpha_j,\lambda_\ell) = \frac{1}{2}\int_\Omega \rho \left|\dot\alpha_i \bm\psi_i + \alpha_j\alpha_k\bm\psi_j\cdot\bm\nabla \bm\psi_k-\nu\alpha_j\bm\nabla^2 \bm\psi_j\right|^2 d\bm{x} - \lambda_\ell D_{\ell j}\dot{\alpha}_j. \]
The first-order necessary condition for optimality with respect to $\dot{\alpha}_i$ is then $\frac{\partial \mathcal{L}}{\partial \dot{\alpha}_i}=0$, which yields
\begin{equation}\label{eq:PMPG_NonDivFree}
\int_\Omega \bm\psi_i \cdot \left(\dot\alpha_j \bm\psi_j + \alpha_j\alpha_k\bm\psi_j\cdot\bm\nabla \bm\psi_k-\nu\alpha_j\bm\nabla^2 \bm\psi_j\right) d\bm{x} -\lambda_\ell D_{\ell i}= 0.
\end{equation}

Thus, when the velocity modes are divergence-free, the classical Galerkin projection and the PMPG yield identical dynamical systems. However, with arbitrary (non-divergence-free) modes, the two formulations may result in different dynamics, depending on the way the incompressibility constraint is discretely imposed in the PMPG formulation. In the mixed Galerkin method, incompressibility is enforced in the weak sense through prescribed pressure modes ($B_{\ell j}\alpha_j=0$). In contrast, the PMPG framework allows more flexibility in imposing the discrete continuity constraints ($D_{\ell j} \dot\alpha_j=0$). Different choices of the functionals $\{\mathcal{E}_\ell\}$ (e.g., pointwise evaluation) lead to different admissible evolution spaces. Consequently, the resulting discrete dynamics may differ from those obtained by mixed Galerkin projection.

Moreover, the PMPG formulation naturally extends to non-modal (nonlinear) parameterization of the velocity field (e.g., neural-network representations):
\begin{equation}\label{eq:NonModal_Representation}
\bm{u}(\bm{x},t) = \mathcal{N}\left( \alpha_i(t),\bm{x} \right),
\end{equation}
where $\alpha_i$ denote the parameters (e.g., weights and biases of the neural network, see \cite{PMPG_PINN_Daqaq,PMPG_PINN_PoF1,Hussam_Unsteady_PINNs_CMAME}). In this setting, the velocity field does not lie in a linear subspace but rather on a nonlinear manifold in function space. Consequently, classical Galerkin projection is not directly applicable. However, the PMPG formulation proceeds intuitively, as follows.

The time derivative of the velocity field is written as
\[
\bm{u}_t = \dot{\alpha}_i \mathcal{N}_{\alpha_i},
\qquad
\mathcal{N}_{\alpha_i} := \frac{\partial \mathcal{N}}{\partial \alpha_i},
\]
so that the PMPG cost becomes
\[ \mathcal{A} (\dot\alpha_i) = \frac{1}{2}\int_\Omega \rho \left|\dot\alpha_i \mathcal{N}_{\alpha_i} + \mathcal{N}\cdot\bm\nabla \mathcal{N}-\nu\bm\nabla^2 \mathcal{N}\right|^2 d\bm{x}. \]
Minimization is performed over $\dot{\alpha}_i$ subject to discrete continuity constraints:
\[ \mathcal{E}_\ell\!\left(\nabla\cdot\bm{u}_t\right) = \mathcal{E}_\ell \left(\nabla\cdot\mathcal{N}_{\alpha_j}\right) \dot\alpha_j = D_{\ell j} \dot\alpha_j = 0. \]
Geometrically, the PMPG performs an orthogonal projection of the free acceleration onto the tangent space of the nonlinear manifold defined by the parameterization (\ref{eq:NonModal_Representation}).

\section{Conceptual Distinctions from Classical Variational Principles}
Because the PMPG is a relatively recent development in the fluid mechanics literature, and because it builds upon Gauss's principle of least constraint---a concept not widely emphasized in standard engineering curricula---it is natural that questions may arise regarding its interpretation and relation to more familiar variational principles. In this section, we clarify several recurring points of confusion surrounding the PMPG.

\subsection{Gauss's Principle and the PMPG are Fundamentally Distinct from Stationary Action}
It is important to emphasize that Gauss's principle of least constraint, and its continuum analogue, the PMPG, are not time-integral principles. They differ fundamentally from Hamilton's principle of stationary (or least) action, in which the cost functional is the time integral of a Lagrangian over an entire trajectory. In contrast, Gauss's principle and the PMPG are instantaneous principles. They select, at each fixed time, the admissible acceleration that minimizes a quadratic cost subject to kinematic constraints. No optimization is performed over trajectories in time, and no time integral appears in the formulation. The minimization is local in time and determines the instantaneous evolution, from which the trajectory subsequently follows via integration.

A related distinction concerns the role of endpoint conditions. In Hamilton's principle, the variational formulation prescribes boundary conditions in time (typically fixing initial and final configurations). In contrast, Gauss's principle and the PMPG require only the instantaneous configuration (position and velocity). They determine the corresponding evolution without reference to a final state. In this sense, the formulation is naturally aligned with forward dynamical evolution and does not presuppose knowledge of future configurations \cite{Tonti}.

Moreover, Hamilton's principle yields a stationarity condition for the action integral, which does not necessarily imply a strict minimum. In contrast, Gauss's principle of least constraint, and the PMPG in the continuum setting, are strict quadratic minimization principles. The associated functional is strongly convex, and the minimizing acceleration is unique. Accordingly, the interpretive statements made in the previous section---namely, that the realized evolution is the one requiring the smallest magnitude of the pressure force---are not teleological. Rather, they follow from strict instantaneous minimality and orthogonal projection in a Hilbert space, not from a time-global optimality condition. In this sense, the reasoning is entirely dynamical and forward in time.

\subsection{The PMPG Does not Follow from the Dirichlet Principle}
Despite its name, the \textit{Principle of Minimum Pressure Gradient} does not involve minimizing the functional
\[ \|\nabla p\|_{\mathbb{L}^2}^2 = \int_\Omega |\bm\nabla p|^2 d\bm{x}\]
over scalar pressure fields. In other words, the PMPG is not a variational principle posed for $p$ itself. Rather, it seeks the acceleration vector field $\bm{u}_t$ that minimizes the cost functional $\mathcal{A}$, defined in Eq. (\ref{eq:PMPG_Cost}). Although the two cost functionals may appear similar---both reflecting the $\mathbb{L}^2$-norm of the pressure gradient---they are equivalent only at the minimizing solution. For a general  kinematically-admissible candidate $\bm{u}_t$, the integrand of the cost $\mathcal{A}$ is not necessarily a gradient field. It becomes a gradient only at the minimizer, as a consequence of orthogonality to the divergence-free subspace. From this viewpoint, the terminology ``Minimum Pressure Gradient" should not be interpreted as a minimization over the space of gradient fields.

Indeed, the two variational formulations---the PMPG and minimizing $\|\nabla p\|_{\mathbb{L}^2}^2$ over $p$---are fundamentally different. Classical potential theory shows, via the Dirichlet principle (e.g., \cite{COV_Dacorogna}), that minimizing $\|\nabla p\|_{\mathbb{L}^2}^2$ over scalar fields $p$ with prescribed Dirichlet boundary data leads to the Laplace equation: $\bm\nabla^2 p=0$. In contrast, Lemma 1 establishes that minimizing the PMPG cost functional $\mathcal{A}$ with respect to the acceleration field $\bm{u}_t$ yields the Navier-Stokes equation. The associated pressure $p$ arises as the Lagrange multiplier enforcing incompressibility. Taking the divergence of the Navier-Stokes equation thus gives the familiar pressure Poisson equation
\begin{equation}\label{eq:Pressure_Poisson}
\bm\nabla^2 p = \rho\nabla\cdot\bm{u}_t^{\rm{free}} = \nabla\cdot\left[-\rho\bm{u}\cdot\bm\nabla \bm{u}+\bm\nabla\cdot\bm{\tau}+\rho\bm{f}\right],
\end{equation}
which is generally inhomogeneous and subject to boundary conditions derived from the velocity field and momentum balance, not from prescribed Dirichlet data on $p$. Hence, the two variational formulations differ not only in their optimization variables ($p$ versus $\bm{u}_t$) but also in their admissible spaces, boundary conditions, and resulting governing equations.

On the other hand, if one begins with the pressure Poisson equation (\ref{eq:Pressure_Poisson}) and applies the Dirichlet principle for Poisson problems, it follows that $p$ minimizes the functional
\[ \int_\Omega \left[\frac{1}{2} |\bm\nabla p|^2 + \rho p \nabla\cdot\bm{u}_t^{\rm{free}} \right] d\bm{x}, \]
which is fundamentally different from the PMPG cost functional $\mathcal{A}$. The above functional is defined over scalar fields $p$
and includes a linear source term involving $\nabla\cdot\bm{u}_t^{\rm free}$ (which is generally non-zero), whereas $\mathcal{A}$ is a strictly convex quadratic functional defined over admissible acceleration fields $\bm{u}_t$.

\subsection{The Instantaneous Nature of the PMPG and Its Relation to Turbulence}
One recurring question regarding the PMPG is its relation to turbulence. For example, one may ask: if the evolution at every instant minimizes the magnitude of the pressure gradient required to enforce incompressibility, how can turbulence develop? In canonical examples such as fully developed laminar channel flow, the steady parabolic profile corresponds to a state in which the instantaneous PMPG cost vanishes; i.e., it cannot decrease any further. From this viewpoint, it may appear that a transition to turbulence contradicts the minimizing philosophy underlying Gauss's principle and the PMPG.

To clarify this point, it is essential emphasize that the mathematical statements presented here do \emph{not} imply that the PMPG cost functional $\mathcal{A}$ is minimized with respect to the velocity field $\bm{u}(\bm{x};t)$ itself. Indeed, if one  were to set the first variation of $\mathcal{A}$ with respect to $\bm{u}(\bm{x};t)$ to zero, the Navier-Stokes equation would \emph{not} be recovered as a necessary condition. Rather, the Navier-Stokes equation arises as the necessary condition for minimizing the PMPG functional $\mathcal{A}$ with respect to the \emph{local acceleration} $\bm{u}_t(\bm{x};t)$, with the instantaneous velocity field $\bm{u}(\bm{x};t)$ held fixed. The variation is therefore taken over spatial fields at a fixed time; and time enters only as a parameter. In this sense, the PMPG is an instantaneous principle governing the admissible accelerations at a given configuration. This interpretation is fully consistent with the philosophy of Gauss's principle as discussed in Sec. \ref{Sec:Gauss} and illustrated through the double-pendulum example.

From this perspective, starting at a given flow field, the PMPG determines the \emph{instantaneous evolution} from that configuration. Hence, if the given configuration is an equilibrium of the Navier--Stokes system (such as the laminar parabolic profile in channel flow), the optimal acceleration is expected to be zero: the best evolution is no evolution. By the two-way equivalence established in Theorem~2, this conclusion necessarily coincides with the Navier--Stokes dynamics. Indeed, if the flow is exactly at an equilibrium configuration, it will remain there under Navier--Stokes dynamics in the absence of perturbations. It is only when the equilibrium is perturbed that the flow transitions into turbulence. Similarly, the PMPG will not result in a non-trivial evolution from the laminar profile $\bm{u}_L$ unless a perturbation $\delta \bm{u}$ is introduced. At the perturbed state $\bm{u} = \bm{u}_L+\delta \bm{u}$, the PMPG cost associated with zero acceleration is generally non-zero:
\[ \mathcal{A}(\bm{u}_t=0;\bm{u}_L+\delta \bm{u})>0 \]
and there may exist admissible evolutions that reduce the instantaneous cost; i.e., there exists $\bm{u}_t^*$ such that
\[ \mathcal{A}(\bm{u}_t^*;\bm{u}_L+\delta \bm{u})< \mathcal{A}(\bm{u}_t=0;\bm{u}_L+\delta \bm{u}). \]
The minimizing acceleration $\bm{u}_t^*$ then determines the subsequent evolution, which may amplify the perturbation if the equilibrium is unstable.

Moreover, while the optimal $\bm{u}_t^*(\bm{x},t)$ at a given instant $t$ minimizes the instantaneous cost for the fixed configuration $\bm{u}(\bm{x},t)$
\[ \mathcal{A}^*(t):=\mathcal{A}\left(\bm{u}_t^*(\bm{x},t);\bm{u}(\bm{x},t)\right)\leq \mathcal{A}\left(\bm{u}_t;\bm{u}(\bm{x},t)\right) \]
over all kinematically admissible evolutions $\bm{u}_t$, it does not necessarily lead to a smaller cost $\mathcal{A}$ at the next time step; i.e.,  $\mathcal{A}^*(t+\Delta t)$ may be larger than $\mathcal{A}^*(t)$. Indeed, evolving the flow field for an infinitesimal time step $\Delta t$ according to the optimal acceleration
\[ \bm{u}(\bm{x},t+\Delta t) = \bm{u}(\bm{x},t) + \bm{u}_t^*(\bm{x},t) \Delta t \]
yields a new configuration at which a new minimization problem is posed. The resulting minimal value
\[ \mathcal{A}^*(t+\Delta t):=\mathcal{A}\left(\bm{u}_t^*(\bm{x},t +\Delta t);\bm{u}(\bm{x},t)+ \bm{u}_t^*(\bm{x},t) \Delta t\right)\]
need not be smaller than $\mathcal{A}^*(t)$.  In other words, minimizing the pressure-gradient norm at each instant does not imply that this norm is a Lyapunov functional or that it decreases monotonically in time. This behavior is entirely consistent with Navier--Stokes dynamics, for which the pressure-gradient norm is not, in general, a monotonically decreasing quantity.

In summary, the two-way equivalence established in Theorem~2 implies that the PMPG provides an alternative formulation of the Navier--Stokes dynamics for smooth flow fields. It recasts the same dynamics in a minimization framework. Consequently, if the Navier--Stokes equations do not generate turbulence from an equilibrium configuration, neither does the PMPG. Conversely, whenever the Navier--Stokes dynamics produces temporal fluctuations in the pressure-gradient norm, the PMPG necessarily reproduces the same behavior.

\section{Relation to the Variational Theory of Lift}
The preceding discussion indicates that there is no \textit{direct} mathematical implication between the PMPG (Lemma~1 or Theorem~2) and the variational theory of lift \cite{Variational_Lift_JFM}. This \textit{theory} was introduced to address the classical closure issue in two-dimensional potential flow. Given a smooth body $\mathcal{B}\subset\mathbb{R}^2$ with exterior fluid domain $\Omega=\mathbb{R}^2\setminus\mathcal{B}$, the steady, incompressible, Euler equation admits a one-parameter family of smooth solutions
\begin{equation}\label{eq:Airfoil_Family}
    \bm{u}(\bm{x};\Gamma) = \bm{u}_0(\bm{x}) + \Gamma \bm{u}_1(\bm{x}), \;\; \Gamma\in\mathbb{R},
\end{equation}
where $\bm{u}_0$ is the non-circulatory flow (i.e., it has zero circulation $\oint_{\partial\mathcal{B}} \bm{u}_0\cdot d\bm{x}=0$); $\bm{u}_1$ is the circulatory flow with unit circulation ($\oint_{\partial\mathcal{B}} \bm{u}_1\cdot d\bm{x}=1$); and the scalar parameter $\Gamma$ therefore represents the total circulation around the body associated with each member of the family (\ref{eq:Airfoil_Family}).

The family (\ref{eq:Airfoil_Family}) satisfies the steady incompressible Euler system for every value of $\Gamma$. Indeed, for every $\Gamma$: (i) It satisfies the steady momentum equation
\[ \rho \bm{u}\cdot\bm\nabla \bm{u} = -\bm\nabla p \]
for some pressure field $p$, since both $\bm{u}_0$, $\bm{u}_1$ are irrotational, and therefore $\bm{u}(\bm{x};\Gamma)$ is irrotational for every $\Gamma$. (ii) It satisfies incompressibility $\bm\nabla \cdot \bm{u}=0$ because both $\bm{u}_0$, $\bm{u}_1$ are divergence-free. (iii) It satisfies the no-penetration boundary condition $\bm{u}\cdot \bm{n}=0$ on $\partial \mathcal{B}$ since both $\bm{u}_0$, $\bm{u}_1$ are tangential to the body surface. (iv) It satisfies the far-field condition $\lim_{|\bm{x}| \to \infty} \bm{u}(\bm{x};\Gamma)=(U,0)$ because $\bm{u}_0$ approaches the prescribed freestream velocity $U$ and $\bm{u}_1$ decays at infinity. Consequently, the family~(\ref{eq:Airfoil_Family}) constitutes a legitimate one-parameter family of steady Euler solutions for arbitrary circulation $\Gamma$.

When the body possesses a sharp trailing edge, all members of the family (\ref{eq:Airfoil_Family}) exhibit singular behavior at that edge, except for a unique member $\bm{u}_K$ that remains bounded everywhere in the domain. This singular structure provides a natural selection criterion within the family: the physically relevant solution is the one that eliminates the trailing-edge singularity. This singularity-removal criterion is known as the Kutta condition \cite{Kutta_Crighton} and has served as the cornerstone of classical airfoil theory for over a century. However, when the body is smooth (i.e., without a sharp edge), there is no generally accepted physical or mathematical criterion that uniquely selects a member from the family. The steady Euler problem therefore remains non-unique in this setting.

Although each member in the family (\ref{eq:Airfoil_Family}) satisfies the Euler system, they generally yield different values of the PMPG cost. In the steady, inviscid setting, the functional reduces to the $\mathbb{L}^2$-norm of the convective acceleration:
\begin{equation}\label{eq:Steady_Appellian}
\mathcal{A}_s(\Gamma) = \frac{1}{2} \int_\Omega |\bm{u}(\bm{x};\Gamma) \cdot \bm\nabla \bm{u}(\bm{x};\Gamma) |^2 d\bm{x}.
\end{equation}
Since each member satisfies the steady Euler equation, it follows that, along this solution family,
\[
\mathcal{A}_s(\Gamma) = \|\nabla p\|_{\mathbb{L}^2}^2(\Gamma).
\]
Thus, each circulation level $\Gamma$ requires a distinct pressure field to enforce the constraints of incompressibility and no-penetration. So, inspired by the philosophy of Gauss's principle, Gonzalez and Taha \cite{Variational_Lift_JFM} proposed a variational closure criterion: Among all admissible steady solutions (\ref{eq:Airfoil_Family}), select the member that minimizes the magnitude of the pressure gradient required to enforce the constraints:
\begin{equation}\label{eq:Kutta_General}
\Gamma^* =\underset{\Gamma}{\rm{argmin}} \;\; \mathcal{A}_s(\Gamma).
\end{equation}

For airfoils with a sharp trailing edge, the minimizing solution $\bm{u}^*$ converges, in the limit of vanishing trailing-edge radius, to Kutta's solution $\bm{u}_K$. In this sense, the classical Kutta condition is recovered as a special case of the variational closure criterion (\ref{eq:Kutta_General}) for sharp-edged geometries. For smooth airfoils, Taha and Gonzalez \cite{Kutta_Flat_Plate} reported quantitative agreement between the minimizing circulation $\Gamma^*$ predicted by the variational framework and that obtained from Reynolds-averaged Navier--Stokes simulations.

Although the variational closure (\ref{eq:Kutta_General}) follows the same philosophy as Gauss's principle and the PMPG, it does not follow directly from the mathematical statements presented in Lemma~1 or Theorem~2. The latter establish minimization with respect to the local acceleration $\bm{u}_t$ (or its finite-dimensional parameters, as discussed in Sec. \ref{Sec:Galerkin}) for a \emph{fixed} velocity field $\bm{u}$. In contrast, the minimization (\ref{eq:Kutta_General}) is carried out with respect to the parameter $\Gamma$, which parameterizes the velocity field $\bm{u}$ itself. Moreover, the minimization is performed over a family of steady Euler solutions. These candidates are not merely kinematically admissible; they already satisfy the dynamical equations of motion. In this sense, the variational criterion (\ref{eq:Kutta_General}) does not select a dynamical solution out of kinematically-admissible ones. Rather, it acts as an \emph{additional} selection criterion within a family of dynamically valid steady solutions, motivated by the same philosophical foundation underlying Gauss's principle and the PMPG.

Having clarified the fundamental distinction between the PMPG and the variational closure~(\ref{eq:Kutta_General}), it is nevertheless conceivable that a deeper, indirect connection exists between the two frameworks, which motivates the investigation of some mathematical questions, presented in the next section. However, before formulating those conjectures, we present another example in which a Gauss-inspired steady minimization reproduces a classical viscous asymptotic result. This example may provide insight into the possible structural link between the instantaneous PMPG dynamics and steady variational selection.

It has been known since the early experiments of Prandtl \cite{prandtl1931hydro} that sufficiently rapid rotation ($\kappa=\frac{\omega b}{U}\gg1$) of a circular cylinder suppresses separation and vortex shedding, leading to an attached flow. Rayleigh's seminal analysis \cite{Rayleigh_Circular_Cylinder} clarified that viscosity transfers the cylinder's rotation into circulation in the outer flow. Outside the boundary layer, the motion becomes irrotational with circulation $\Gamma$.

The central question becomes: what is the value of $\Gamma$ generated by a cylinder of radius $b$ rotating with angular velocity $\omega$ in a free stream $U$? As Rayleigh remarked, ``\textit{friction is the immediate cause of the whirlpool motion},'' indicating that circulation cannot be determined without accounting for viscous effects. The difficulty extends beyond the absence of a Kutta-like condition: in an inviscid formulation, the no-penetration boundary condition specifies only the normal velocity. The tangential surface velocity induced by rotation cannot be imposed. Thus, for a rotating cylinder, where the boundary motion is purely tangential, the ideal-fluid formulation is insensitive to the rotation. As noted explicitly by Moore \cite{moore1957flow}, ``\textit{it is not possible to determine $\Gamma$ without solving the Navier–Stokes equations}.''

Glauert \cite{Glauert_Rotating_Cylinder} provided the first comprehensive theoretical treatment. He modeled the outer flow as irrotational with unknown circulation; i.e., similar to the family (\ref{eq:Airfoil_Family}). To determine the value of this circulation, he solved Prandtl's boundary-layer equations under the no-slip condition. Exploiting the large-$\kappa$ assumption and expanding in $\epsilon=1/\kappa$, he obtained
\begin{equation}\label{eq:Glauert_Result}
\frac{\Gamma}{\Gamma_\omega}
= 1-\frac{3}{\kappa^2}
+O\!\left(\frac{1}{\kappa^4}\right),
\end{equation}
where $\Gamma_\omega=2\pi b^2\omega$ is the circulation associated with the cylinder's surface motion. The ratio $\Gamma/\Gamma_\omega$ may be interpreted as the transfer efficiency of circulation from the rotating boundary layer to the outer flow. The expansion (\ref{eq:Glauert_Result}) reveals a well-defined inviscid limit:
\[ \frac{\Gamma}{\Gamma_\omega} = 1-\frac{3}{\kappa^2}. \]

Shorbagy and Taha \cite{Shorbagy_Magnus_AIAA} proposed a Gauss-inspired variational closure that recovers this inviscid limit without solving the nonlinear boundary-layer equations. They followed Batchelor's two-stage description \cite{Batchelor} of the problem ``\textit{in  two stages. First the cylinder is given a steady angular velocity $\omega$ in fluid initially at rest; ... vorticity is generated in the fluid and diffuses to infinity, leaving a steady irrotational motion with circulation $2\pi b^2 \omega$. The cylinder is then given a translational velocity $U$}." In the language of Gauss, they treated the irrotational flow induced by $\Gamma_\omega$ as the impressed (free) motion. This free motion has known acceleration
\[ \bm{a}^{\rm{free}}=\frac{ \Gamma_\omega^2}{4\pi^2 r^3}\bm{e}_r. \]

When a uniform stream $U$ is superposed, the actual outer flow is selected, in the spirit of Gauss's principle, as the member of the family (\ref{eq:Airfoil_Family}) that deviates least from this impressed motion. Accordingly, the Gaussian deviation cost
\begin{equation}\label{eq:Cylinder_Cost}
\mathcal{A}_{s,\rm{rot}}(\Gamma) = \frac{1}{2}\rho \int_\Omega \left[\bm{u}(\bm{x};\Gamma)\cdot\bm\nabla \bm{u}(\bm{x};\Gamma)-\bm{a}^{\rm{free}}\right]^2 d\bm{x}
\end{equation}
is minimized with respect to $\Gamma$.

\begin{wrapfigure}{l}{0.50\textwidth}
\vspace{-0.1in}
 \begin{center} \vspace{-0.1in}
 \includegraphics[width=7cm]{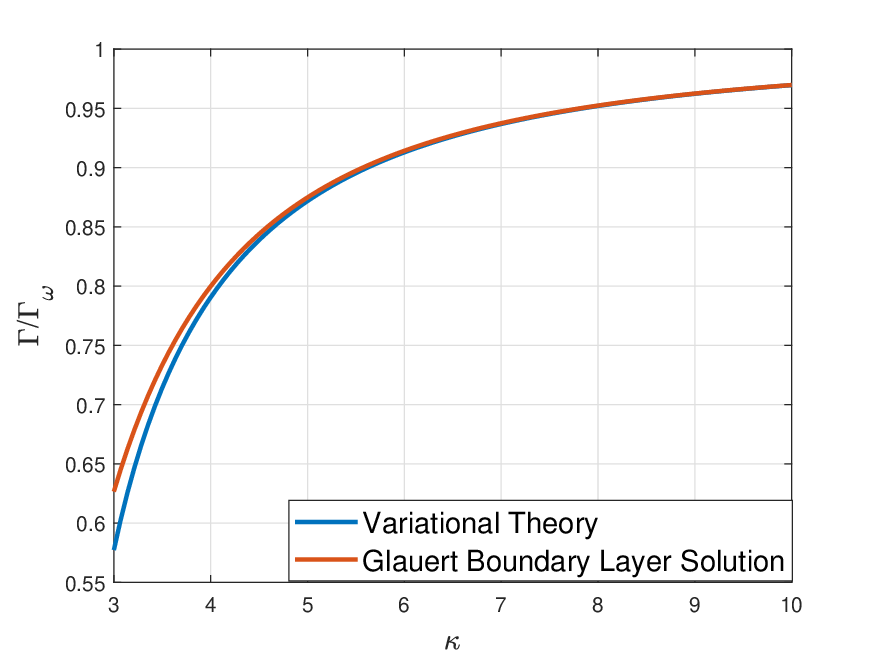}  \vspace{-0.2in}
 \caption{Comparison for the variation of the normalized circulation $\frac{\Gamma}{\Gamma_\omega}$ with the normalized rotational speed $\kappa=\frac{\omega b}{U}$ between Glauert's viscous boundary layer solution (\ref{eq:Glauert_Result}) and the PMPG solution (\ref{eq:PMPG_Rotating_Cylinder}).}
 \label{Fig:Magnus_Result} \vspace{-0.2in}
 \end{center}\vspace{-0.3in}
\end{wrapfigure} \noindent The necessary condition for minimization $\frac{d \mathcal{A}_{s,\rm{rot}}}{d \Gamma}=0$ yields:
\begin{equation}\label{eq:PMPG_Rotating_Cylinder}
\frac{\Gamma^*}{\Gamma_\omega} = \sqrt{1-\frac{6}{\kappa^2}}.
\end{equation}

 Figure~\ref{Fig:Magnus_Result} compares (\ref{eq:PMPG_Rotating_Cylinder}) with Glauert's asymptotic result (\ref{eq:Glauert_Result}). Expanding (\ref{eq:PMPG_Rotating_Cylinder}) for large $\kappa$ gives
\[ \frac{\Gamma^*}{\Gamma_\omega} =  1-\frac{3}{\kappa^2}+O\left(\frac{1}{\kappa^4}\right),\]
which coincides with Glauert's inviscid limit to second order. Thus, a one-dimensional variational minimization reproduces the asymptotic circulation obtained from the viscous boundary-layer analysis.

\section{Dynamical Minimization Versus Steady Optimality}
The results obtained from variational closure, discussed above, motivate the following mathematical question: since the flow evolves at every instant by minimizing the cost $\mathcal{A}(\bm{u}_t;\bm{u})$, suppose that the evolution converges to a steady state equilibrium $\bar{\bm{u}}$. Does it necessarily imply that $\bar{\bm{u}}$ minimizes the steady cost $\mathcal{A}_s(\bm{u}):=\mathcal{A}(\bm{0};\bm{u})$? For a general dynamical system, the answer is negative. The following example provides a counterexample.

\begin{wrapfigure}{l}{0.50\textwidth}
  \begin{center}
  \includegraphics[width=7cm]{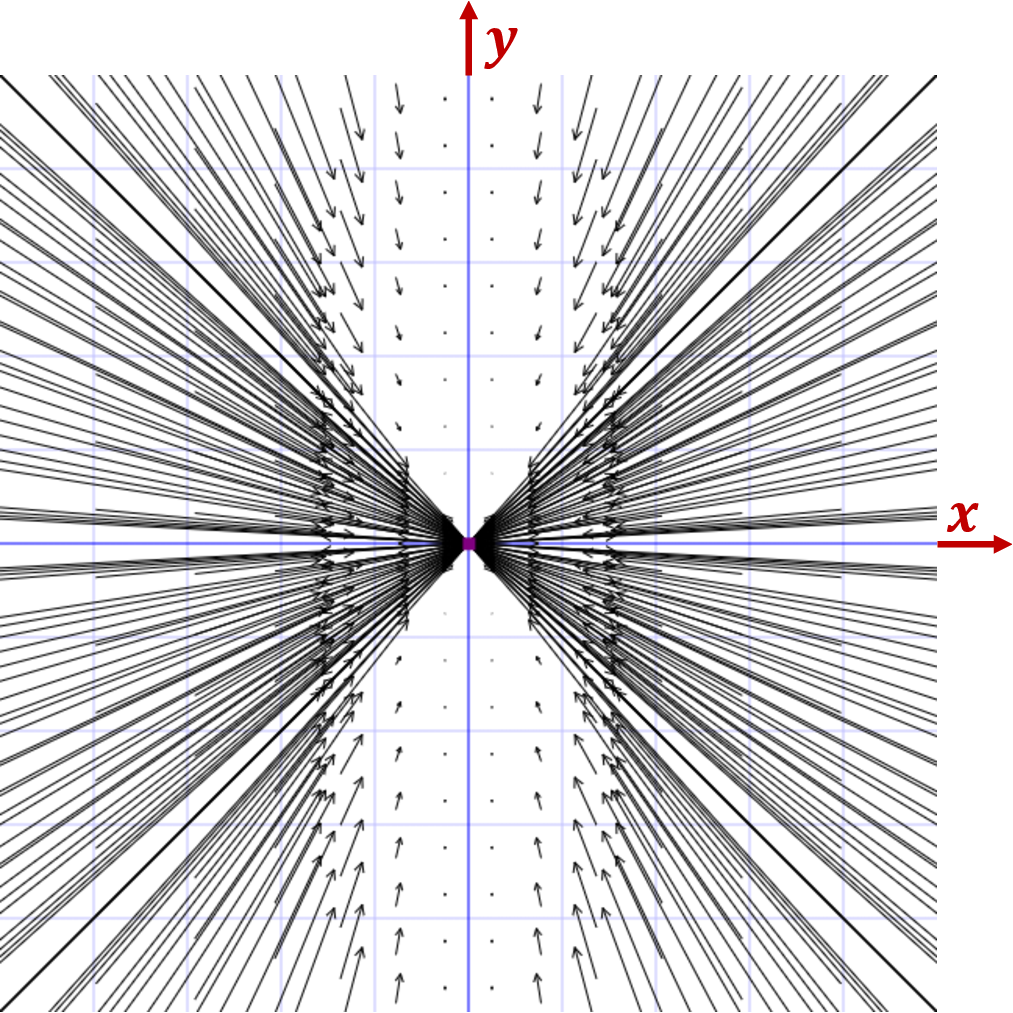}
  \caption{Phase portrait of the two-dimensional dynamical system (\ref{eq:Exmple_Dynamical_Sys}). It has the entire $y$-axis ($x=0$) as a line of equilibria. However, only the origin is stable; starting from any point in the plane (except on the $y$-axis), the system converges to the origin.}\vspace{-0.1in}
  \label{Fig:Phase_Plane_Example}
  \end{center}
 \end{wrapfigure}

Consider the two-dimensional dynamical system with state variables $\bm\chi=(x,y)$, whose instantaneous evolution $\dot{\bm\chi}$ is determined by minimizing the cost
\[ \mathcal{S}(\dot{\bm\chi};\bm\chi) = \left(\dot{x}+x^3\right)^2 + \left(\dot{y}+yx^2\right)^2 + \left(y-3\right)^2.\]
This minimization results in the dynamical system
\begin{equation}\label{eq:Exmple_Dynamical_Sys}
 \dot{x} = -x^3 \;\;\mbox{and} \;\; \dot{y}=-yx^2,
 \end{equation}
which has the entire $y$-axis ($x=0$) as a continuous family of equilibrium points. However, only the origin is stable; starting from any initial condition with $x\neq0$, the trajectory converges to the origin, as illustrated in the phase portrait shown in Fig. \ref{Fig:Phase_Plane_Example}.

 \noindent In this example, the steady cost is given by
\[ \mathcal{S}_s(\bm\chi):=\mathcal{S}(\bm{0};\bm\chi) = x^6 + y^2 x^4 +\left(y-3\right)^2,  \]
which is minimum at $(0,3)$. In contrast, the stable equilibrium point selected by the dynamics (i.e., the origin) is not a local minimizer of the steady cost.

This example was devised to mirror, in a simplified setting, the structure of the potential-flow lift problem. Similar to the $y$-axis in this system, ideal-fluid dynamics admits a continuous family of steady solutions, parameterized by the $\Gamma$-axis. Yet, in physical reality, only one member of this family is realized; for a sharp-edged airfoil, this is Kutta's circulation. The variational theory of lift asserts that this physically realized equilibrium corresponds to the minimizer of the associated steady cost. The counterexample above shows that such a conclusion does not hold for general dynamical systems: instantaneous minimization of a cost does not, in general, imply that the dynamically selected equilibrium minimizes the corresponding steady cost. Nevertheless, one must exercise caution in drawing parallels. The two-dimensional dynamical system introduced above, although structurally analogous to the lift-selection problem, is fundamentally different in nature; it is not derived from a mechanical or continuum framework. The analogy is therefore suggestive, but not conclusive.

If additional structural properties are imposed on the dynamics, the implication that a stable equilibrium $\bar{\bm{u}}$ minimizes the steady cost $\mathcal{A}_s(\bm{u})$ can be ensured. For instance, suppose that in a neighborhood of a stationary solution $\bar{\bm{u}}$, the pressure-gradient norm (equivalently, the PMPG cost) is monotonically non-increasing under the dynamics. Then $\bar{\bm{u}}$ must necessarily be a local minimizer of the steady cost $\mathcal{A}_s$. The following lemma formalizes this observation.\\

\textbf{Lemma 3}: Consider a dynamical system with state variables $\bm\chi \in \mathcal{X}$ that evolves at every instant by minimizing a cost $\mathcal{S}(\dot{\bm\chi};\bm\chi)$, which depends smoothly on the instantaneous evolution $\dot{\bm\chi}$ and the current configuration $\bm\chi$. That is, the dynamics is defined according to the unique optimal evolution
\[ \dot{\bm\chi}^* (\bm\chi):=  \underset{\dot{\bm\chi}}{\rm{argmin}} \; \mathcal{S}(\dot{\bm\chi};\bm\chi).\]
At each configuration $\bm\chi$, the optimal cost is then given by 
\[ \mathcal{S}^* (\bm\chi) := \mathcal{S}(\dot{\bm\chi}^* (\bm\chi) ;\bm\chi). \]

Assume that $\bar{\bm\chi}$ is an asymptotically stable equilibrium of the induced dynamics such that there exists a neighborhood $\mathcal{N}\subset\mathcal{X}$ of $\bar{\bm\chi}$ where, along trajectories of the system, 
\[ \frac{d \mathcal{S}^*}{dt} (\bm\chi(t)) \leq0 \;\; \forall \; \bm\chi(t) \in \mathcal{N}.\]

Then, $\bar{\bm\chi}$ must be a local minimizer of the steady cost $\mathcal{S}_s(\bm\chi):=\mathcal{S}(\bm{0};\bm\chi)$. In particular, we have
\[ \mathcal{S}_s(\bar{\bm\chi}) \leq \mathcal{S}_s(\bm\chi) \;\; \forall \; \bm\chi \in \mathcal{N}.\]

\noindent\textbf{Proof:} Since $\bar{\bm\chi}$ is asymptotically stable and $\mathcal{S}^*$ is continuous in $\bm\chi$ (which follows from the assumed smoothness of $\mathcal{S}$ in both arguments and uniqueness of the minimizer), we have
\[ \lim_{t\to\infty} \bm\chi(t) = \bar{\bm\chi} \; \Rightarrow \; \lim_{t\to\infty} \mathcal{S}^*(\bm\chi(t)) = \mathcal{S}^*(\bar{\bm\chi}). \]
Since $\bar{\bm\chi}$ is asymptotically stable, we may restrict $\mathcal{N}$, if necessary, so that it is contained in the region of attraction of $\bar{\bm\chi}$; i.e., trajectories starting at any $\bm\chi \in \mathcal{N}$ converge to $\bar{\bm\chi}$. Moreover, by assumption, $\mathcal{S}^*$ is non-increasing along trajectories in $\mathcal{N}$. Thus, for any initial condition $\bm\chi \in \mathcal{N}$,
\[ \mathcal{S}^*(\bar{\bm\chi}) = \lim_{t\to\infty}\mathcal{S}^*(\bm\chi(t)) \le \mathcal{S}^*(\bm\chi). \]
Equivalently,
\[ \mathcal{S}(\dot{\bm\chi}^* (\bar{\bm\chi}) ;\bar{\bm\chi}) \leq \mathcal{S}(\dot{\bm\chi}^* (\bm\chi) ;\bm\chi)  \;\; \forall \; \bm\chi \in \mathcal{N}.\]
Since $\bar{\bm\chi}$ is an equilibrium, its optimal evolution vanishes: $\dot{\bm\chi}^*(\bar{\bm\chi})=\bm{0}$. Hence,
\[ \mathcal{S}(\bm{0} ;\bar{\bm\chi}) \leq \mathcal{S}(\dot{\bm\chi}^* (\bm\chi) ;\bm\chi) \;\; \forall \; \bm\chi \in \mathcal{N}.\]
Finally, by definition of optimality, we have $\mathcal{S}(\dot{\bm\chi}^* (\bm\chi) ;\bm\chi) \leq \mathcal{S}(\bm{0} ;\bm\chi)$, which yields
\[ \mathcal{S}(\bm{0};\bar{\bm\chi}) \leq \mathcal{S}(\bm{0} ;\bm\chi) \;\; \forall \; \bm\chi \in \mathcal{N}.\]
Equivalently,
\[ \mathcal{S}_s(\bar{\bm\chi}) \leq \mathcal{S}_s(\bm\chi) \;\; \forall \; \bm\chi \in \mathcal{N},\]
which proves that $\bar{\bm\chi}$ is a local minimizer of the steady cost $\mathcal{S}_s$. \hfill $\blacksquare$

The above Lemma establishes that, if the $\mathbb{L}^2$-norm of the pressure gradient (equivalently, the optimal PMPG cost $\mathcal{A}^*$) is non-increasing along the flow dynamics in a neighborhood of a steady solution, then that steady solution must be a local minimizer of the steady PMPG cost
\begin{equation}\label{eq:Steady_PMPG_Cost}
\mathcal{A}_s(\bm{u}) = \frac{1}{2} \int_\Omega |\bm{u} \cdot \bm\nabla \bm{u} - \nu\bm\nabla^2 \bm{u} |^2 d\bm{x}.
\end{equation}
This behavior is observed in \textit{monotonically stable} flows at sufficiently low Reynolds numbers, where the perturbation kinetic energy decays monotonically toward the steady state, without overshoot or non-modal transient growth \cite{Schmid_Stability_Book}. This observation provides a possible explanation for why minimization of the steady cost $\mathcal{A}_s$ can successfully recover steady solutions in certain cases, such as the lid-driven cavity at low Reynolds numbers \cite{PMPG_PINN_Daqaq}. However, extending this reasoning beyond such regimes requires further mathematical analysis.

Another observation supporting the hypothesis that stable steady solutions may minimize the steady cost $\mathcal{A}_s$ arises from a time-discrete formulation of the PMPG framework. Discretizing the instantaneous minimization in time with a forward Euler step yields the problem
\begin{equation}\label{eq:PMPG_Cost_Explicit}
\min_{\bm{u}_{k+1}} \;\; \mathcal{A} (\bm{u}_{k+1}) = \frac{1}{2}\int_\Omega \rho \left|\frac{\bm{u}_{k+1} - \bm{u}_{k}}{\Delta t} + \bm{u}_k\cdot\bm\nabla \bm{u}_k-\nu\bm\nabla^2 \bm{u}_k\right|^2 d\bm{x}
 \end{equation}
over the admissible set 
\begin{equation}\label{eq:Admissible_Set_Discrete_Time}
\mathcal{U} =\{\bm{u}\in\mathbb{L}^2| \; \bm\nabla \cdot\bm{u}=0 \; \forall \; \bm{x}\in\Omega, \;\; \mbox{and} \;\; \bm{u}\cdot \bm{n} = h(\bm{x})\; \forall \; \bm{x}\in\partial\Omega\}.
 \end{equation}
Under the same convexity and admissibility assumptions discussed earlier, following the reasoning of Theorem 2, one can show a two-way equivalence between this discrete-time \textit{explicit} minimization problem and the corresponding Navier-Stokes system:
\[ \bm{u}_{k+1} = \bm{u}_{k} - \Delta t \left[ \bm{u}_k\cdot\bm\nabla \bm{u}_k-\nu\bm\nabla^2 \bm{u}_k +\bm\nabla p_k \right], \;\; \bm\nabla\cdot \bm{u}_{k+1} = 0.\]
Thus, at each time step, advancing the Navier–Stokes solution is equivalent to solving the instantaneous discrete PMPG minimization problem (\ref{eq:Admissible_Set_Discrete_Time}).

A standard strategy in computational practice for obtaining steady-state solutions is to employ \textit{implicit} time integrators with large time steps \cite{Hirsch_Book}. Heuristically, if the time step $\Delta t$ is sufficiently large, a single implicit advancement may drive the solution toward a stable steady state $\bar{\bm{u}}$. Writing the implicit analogue of the discrete PMPG minimization problem (\ref{eq:PMPG_Cost_Explicit})
\begin{equation}\label{eq:PMPG_Cost_Implicit}
 \min_{\bm{u}_{k+1}} \;\; \mathcal{A} (\bm{u}_{k+1}) = \frac{1}{2}\int_\Omega \rho \left|\frac{\bm{u}_{k+1} - \bm{u}_{k}}{\Delta t} + \bm{u}_{k+1}\cdot\bm\nabla \bm{u}_{k+1}-\nu\bm\nabla^2 \bm{u}_{k+1}\right|^2 d\bm{x},
  \end{equation}
and taking the limit $\Delta t \to\infty$, one obtains the steady-state minimization problem:
\[ \min_{\bar{\bm{u}}} \;\; \mathcal{A} (\bar{\bm{u}}) = \frac{1}{2}\int_\Omega \rho \left|\bar{\bm{u}}\cdot\bm\nabla \bar{\bm{u}}-\nu\bm\nabla^2 \bar{\bm{u}}\right|^2 d\bm{x}\equiv \mathcal{A}_s (\bar{\bm{u}}). \]

That is, advancing the flow field using the implicit PMPG formulation (\ref{eq:PMPG_Cost_Implicit}) and taking the large–$\Delta t$ limit reduces the evolution to a direct minimization of the steady functional $\mathcal{A}_s$. Consequently, if the implicit PMPG dynamics converges to a steady-state solution $\bar{\bm{u}}$, then $\bar{\bm{u}}$ must be a local minimizer of the steady cost $\mathcal{A}_s$ over the admissible class.

\section{Mathematical Conjectures}
In the two examples discussed earlier (the airfoil and the rotating cylinder), the minimization of the inviscid version of the steady cost $\mathcal{A}_s$ (e.g., the $\mathbb{L}^2$-norm of the convective acceleration) successfully selects the physically relevant solution from the infinitely many members of Euler's family. This observation suggests that such minimization may provide a principled selection criterion for resolving the non-uniqueness inherent in steady Euler dynamics. To elevate this idea from heuristic observation to mathematical conjecture, one must first clarify what is meant by a \emph{physical} solution of the Euler equations. From a mathematical standpoint, a steady Euler solution may be regarded as physically relevant if it satisfies at least one of the following properties: (i) it is a stable solution; or (ii) it arises as the inviscid limit of a Navier–Stokes solution, and therefore represents the limiting behavior of physically realizable viscous flows at sufficiently high Reynolds numbers.

That is, Kutta's solution for the flow over a sharp-edged airfoil may be regarded as physical because it corresponds either to a stable steady configuration, to the inviscid limit of the Navier-Stokes solution, or to both. The fact that minimization of $\mathcal{A}_s$ recovers Kutta's solution as a special case suggests that this variational selection criterion may represent a necessary condition for (i) stability of steady Euler solutions, or (ii) identification of the inviscid limit of the Navier-Stokes equation within Euler's family. It is noteworthy that, in the rotating-cylinder problem, minimization of $\mathcal{A}_s$ reproduces precisely the inviscid limit obtained from the viscous boundary-layer analysis. This result further supports the possibility that minimization of the convective-acceleration norm encodes structural information traditionally derived from viscous dynamics.

These observations motivate the following two conjecture drafts, proposed as directions for future mathematical investigation. The qualifier \textit{drafts} is used deliberately to emphasize that the statements below are preliminary and not yet expressed in fully rigorous terms. They are offered in the hope that mathematical analysts may refine and formalize them into precise and verifiable conjectures.\\

\textbf{Conjecture Draft 1 (A Necessary Condition for Stability)}: \textit{A stable stationary solution of the incompressible Euler equation must locally minimize the $\mathbb{L}^2$-norm of the convective acceleration}. More precisely, if $\bar{\bm{u}}\in C^k (\Omega)$, $k>1$ is (i) divergence-free $\bm\nabla\cdot\bar{\bm{u}}=0$, (ii) satisfies the no-penetration boundary condition $\bar{\bm{u}}\cdot\bm{n}=h(\bm{x})$ on a smooth boundary $\partial\Omega$, for some prescribed smooth function $h$, (iii) $\bar{\bm{u}}$ goes to a constant at infinity, and (iv) is a \textit{stable stationary solution of Euler's} equation, then it must be a local minimizer of the functional
\[ \mathcal{A}_s (\bm{u})= \frac{1}{2}\int \left|\bm{u}\cdot\bm\nabla \bm{u} \right|^2 d\bm{x}\]
over the family of admissible stationary solutions
\[ \Phi=\left\{ \bm{u}\in C^k (\Omega)|\; \bm{u}\cdot\nabla\bm{u}\in\mathbb{L}^2, \; \bm\nabla\cdot\bm{u}=0, \; \bm{u}\cdot\bm{n}=h(\bm{x}), \; \lim_{r\to\infty}\bm{u} \to \rm{constant}. \right\}\]

The above conjecture draft is presented for the inviscid case. A corresponding investigation for viscous flows, with the steady cost defined by Eq. (\ref{eq:Steady_PMPG_Cost}) is a natural extension and appears equally worthy of rigorous analysis. If the inviscid version were to be established, it would suggest a necessary condition for stability expressed purely in terms of the convective-acceleration norm---a perspective that differs fundamentally from classical energy-based criteria. Such a result would provide a new structural lens on inviscid stability distinct from classical energy–Casimir or eigenvalue-based criteria, though it would not apply to configurations such as pure shear flows, for which the convective acceleration vanishes identically.

The rotating-cylinder example suggests that minimization of the steady functional $\mathcal{A}_s$ may encode information about the inviscid limit of viscous flows.  In particular, the variational closure recovers the asymptotic circulation obtained from the Navier–Stokes boundary-layer analysis. This observation raises the possibility that steady minimization of $\mathcal{A}_s$ may act as a selection criterion for identifying, within Euler's family, the member corresponding to the inviscid limit of Navier–Stokes solutions. This possibility is presented in the following conjectural statement.\\

\textbf{Conjecture Draft 2 (Irrotational Inviscid Limit)}: Let $\bar{\bm{u}}^\nu$ be a smooth steady solution of the two-dimensional incompressible Navier–Stokes equations around a smooth stationary body $\mathcal{B}$, with $\bar{\bm{u}}^\nu \to (U,0)$ at infinity. Assume that, for sufficiently small viscosity $\nu$, the viscous boundary layer remains confined to a thin neighborhood of $\partial\mathcal{B}$, and that outside this boundary layer the flow is well-approximated by an irrotational field.

Let $\mathcal{U}_{\mathrm{outer}}$ denote the family of admissible irrotational flows around the body, typically expressed in the form (\ref{eq:Airfoil_Family}). Suppose that, away from the boundary layer, $\bar{\bm{u}}^\nu$ converges (in an appropriate sense) to an irrotational flow $\bar{\bm{u}}^0$ as $\nu \to 0$. Then $\bar{\bm{u}}^0$ must be a local minimizer of the steady inviscid cost $\mathcal{A}_s= \frac{1}{2} \int_\Omega |\bm{u}\cdot\bm\nabla \bm{u}|^2 d\bm{x}$ over the family $\mathcal{U}_{\mathrm{outer}}$. In particular, the inviscid limit takes the form
\[ \bar{\bm{u}}^0 = \bm{u}_0 + \Gamma^* \bm{u}_1, \]
where $\Gamma^*$ is the minimizer of $\mathcal{A}s$  in Eq. (\ref{eq:Steady_Appellian}) over $\mathcal{U}{\mathrm{outer}}$.

Finally, we must emphasize that the two-way equivalence established by Theorem 2 relies on uniqueness of sufficiently smooth solutions of Navier-Stokes and Euler equations. In regimes where only weak or rough solutions are available and uniqueness is forfeited, the PMPG minimization problem may no longer be well-defined in its classical form: the integrand may fail to lie in $\mathbb{L}^2$, and the time derivative $\bm{u}_t$ may not be defined pointwise. Even if the minimization problem were reformulated in a weaker functional setting---such as restricting to flows that are absolutely continuous in time and redefining the cost in a weaker norm (e.g., $H^{-1}$)---the two-way equivalence need not persist. The Navier–Stokes or Euler equations may admit multiple admissible solutions, whereas the associated minimization problem could produce a unique evolution. In such a situation, it becomes natural to ask whether a suitably defined minimization framework might serve as a selection principle for identifying physically relevant solutions.

This question is closely related to the longstanding problem of convergence of Navier–Stokes solutions to Euler solutions as viscosity vanishes \cite{Masmoudi_Inviscid_Limit1,Majda_Vorticity_Book,Masmoudi_Inviscid_Limit2}. As stated by Eyink \cite{Eyink_PRX}, ``\textit{under very modest assumptions, the viscous Navier-Stokes solution must tend in the limit $Re\to\infty$ to a weak solution of the incompressible Euler equations."} However, weak solutions of Euler are generally non-unique \cite{Euler_Weak_Nonuniqueness}. So, a selection criterion is needed \cite{Eyink_JFM2024}. Among the family of weak solutions of Euler's equation, which solution arises as the zero-viscosity limit? The formulation of a minimization principle capable of providing such a selection mechanism constitutes a challenging problem in mathematical fluid dynamics. The present work does not resolve this question; rather, it merely points to a potential cost functional for such a selection principle: a suitable norm of the pressure force, required to enforce incompressibility, following the philosophy of Gauss.

\section{Conclusion}
In this paper, we present a mathematical analysis of the principle of minimum pressure gradient (PMPG). First, we establish two-way equivalence between the PMPG and the incompressible Navier–Stokes equations (INSE). We prove that a candidate smooth flow field is a solution of the INSE \textit{if and only if} its instantaneous evolution minimizes, at every instant, the $\mathbb{L}^2$-norm of the pressure force required to ensure incompressibility. We show that the PMPG is precisely the minimization formulation of the Leray-Helmholtz projection onto the space of divergence-free vector fields. Any admissible instantaneous evolution (e.g., onset of separation or transition) resulting from the INSE must necessarily minimize the PMPG cost. Conversely, any other kinematically admissible evolution, requiring a strictly larger pressure force to ensure the same constraints, fails to satisfy the INSE. This two-way equivalence shows that the PMPG presents a minimization perspective through which intricate incompressible flow behaviors may be interpreted.

In a finite-dimensional setting with divergence-free modes, we show that the PMPG yields the same dynamics as classical Galerkin projection. However, when non-divergence-free modes are adopted, the mixed Galerkin formulation and the PMPG may produce different dynamical systems. Moreover, the PMPG provides a natural generalization of classical Galerkin projection beyond linear modal expansions, accommodating nonlinear and non-modal representations.

We then examine the relation between instantaneous dynamical minimization and steady variational selection, including its connection to the variational theory of lift. The central question is: since the flow evolves at every instant by minimizing the PMPG cost $\mathcal{A}(\bm{u}_t;\bm{u})$, does convergence to a steady equilibrium $\bar{\bm{u}}$ imply that $\bar{\bm{u}}$ minimizes the steady cost $\mathcal{A}_s(\bm{u}):=\mathcal{A}(\bm{0};\bm{u})$? We show that the answer is negative for a general dynamical system by providing a counterexample. However, if additional structural properties are imposed (e.g., monotonic non-increase of the pressure-gradient norm in a neighborhood of $\bar{\bm{u}}$), the conclusion follows.

We also present a discrete-time formulation of the PMPG. In particular, we show that if the implicit discrete-time PMPG scheme converges to a steady equilibrium, then the steady limit must minimize the steady cost $\mathcal{A}_s$ (the $\mathbb{L}^2$-norm of the pressure gradient). Motivated by these observations, we formulate two conjectures that warrant further mathematical investigation: (i) whether a stable equilibrium necessarily minimizes the steady cost $\mathcal{A}_s$, and (ii) whether the irrotational inviscid limit of Navier-Stokes solutions over a two-dimensional smooth body minimizes $\mathcal{A}_s$ (i.e., the $\mathbb{L}^2$-norm of the convective acceleration in the inviscid setting).

\section*{Acknowledgments}
The author is grateful to the fruitful discussions with Professors Karthik Duraisamy, Neelesh Patankar, Jorn Loviscach, Antony Jameson, Ahmed Roman, and Scott Bollt. The author would also like to acknowledge the support of the National Science Foundation grant number CBET-2332556.

\bibliography{Fluid_Dynamics_References,Aeronautical_Engineering_References,Geometric_Control_References,Dynamics_Control_References,Math_References,My_Published_References,Other_Refs}
\bibliographystyle{unsrt}

\end{document}